\documentclass[%
 preprint,
 amsmath,amssymb,
 aps,
]{revtex4-1}
\usepackage{graphicx}% Include figure files
\usepackage{dcolumn}% Align table columns on decimal point
\usepackage{bm}% bold math
\usepackage{float}

\begin{document}

\title{Exact intermittent solutions in a turbulence multi branch shell model }
\author{Ben Ajzner }
\affiliation{Laboratoire de Physique de l'Ecole normale sup{\'e}rieure, ENS, Universit{\'e} PSL, CNRS, Sorbonne Universit{\'e}, Universit{\'e} de Paris, F-75005 Paris, France}
\author{Alexandros Alexakis}
\affiliation{Laboratoire de Physique de l'Ecole normale sup{\'e}rieure, ENS, Universit{\'e} PSL, CNRS, Sorbonne Universit{\'e}, Universit{\'e} de Paris, F-75005 Paris, France}

% Contact information of the corresponding author
%\corres{Correspondence: alexakis@phys.ens.fr}
%\conference{} % An extended version of a conference paper
% Abstract (Do not insert blank lines, i.e. \\) 
\begin{abstract}
\centerline{ \bf For the commemorative Issue Dedicated to the Memory of Jackson Rea Herring}  %\\
%-
%\\
Reproducing complex phenomena with simple models marks our understanding of the phenomena themselves
and this is what Jack Herring's work demonstrated multiple times. 
In that spirit,
this work studies a turbulence shell model consisting of a hierarchy of structures of different
scales $\ell_n$ such that each structure transfers its energy to two substructures of scale $\ell_{n+1}=\ell_n/\lambda$. 
For this model we construct exact inertial range solutions that display intermittency {\it ie} absence of self-similarity. 
Using a large ensemble of these solutions we investigate how the probability distributions of the velocity modes
change with scale. It is demonstrated that while velocity amplitudes are not scale invariant their ratios are.
Furthermore using large deviation theory we show how  the probability distributions of the velocity modes
can be re-scaled to collapse in a scale independent form.    
Finally, we discuss the implications the present results have for real turbulent flows. %}
\end{abstract}
% Keywords
%\keyword{Turbulence, intermittency} 

% The fields PACS, MSC, and JEL may be left empty or commented out if not applicable
%\PACS{J0101}
%\MSC{}
%\JEL{}

%%%%%%%%%%%%%%%%%%%%%%%%%%%%%%%%%%%%%%%%%%
%\begin{document}
\maketitle

%%%%%%%%%%%%%%%%%%%%%%%%%%%%%%%%%%%%%%%%%%
\section{Introduction}

Constructing simple models that reproduce the phenomenological complex behavior of fluid flows has always been a driving force in turbulence research and is a direction in which Jack Herring's work excelled. 
There are numerous works in his career explaining complex phenomena in fluid dynamics with simplified models
\cite{herring1965self,herring1974approach,herring1975theory,herring1974decay,herring1977statistical,herring1979test,herring1980statistical,herring1980theoretical,herring1982comparative,herring1985comparison,herring1989non,herring1979subgrid,herring1989numerical}. In particular the  energy cascade in scale space is a phenomenon that has met various modeling 
approaches in the literature as 
the direct interaction approximation \cite{kraichnan1959structure,kraichnan1964decay,herring1965self,herring1966self,herring1977statistical},
eddy damping quasi-normal Markovian models,
\cite{orszag1970analytical,leith1971atmospheric,fouquet1975evolution,lesieur20003d}
energy diffusion models \cite{leith1967diffusion,leith1968diffusion} and shell models \cite{desnianskii1974simulation,gledzer1973eb,yamada1987lyapunov,l1998improved}.
Such models have lead to predictions about the direction of cascade, the power-law exponents of
the energy spectra and intermittency. Intermittency that still escapes a firm quantitative understanding  
manifests itself as a deviation from self-similarity and from the prediction obtained on
purely dimensional grounds. 
%
%In this spirit we here construct and study a shell model a binary tree shell model of turbulence that displays intermittency.
In particular, shell models have been used to study  intermittency for many years now. Recent reviews can be found in \cite{biferale2003shell,alexakis2018cascades}. Typically, shell models quantify all structures of a given scale $\ell$
by a single real or complex amplitude $U_\ell$. As such, spatial intermittency that is linked to the appearance of 
rare but extremely intense structures, can not be captured this way. Nonetheless,
the temporal variation of the modes $U_l$ does display intermittency as has been demonstrated by many models  \cite{l1998improved,gledzer1973eb,yamada1987lyapunov,jensen1991intermittency}.  
This type of intermittency has been recently linked to the fluctuation dissipation theorem \cite{aumaitre2023}.
Furthermore a solvable (but not energy conserving) model was also derived and studied in \cite{mailybaev2021solvable}. 

In the spirit discussed in the first paragraph we here construct and study a binary tree shell model for turbulence that 
displays intermittency. In this model energy at each scale is split between multiple different structures.
Each structure transfers its energy into two smaller scale structures of smaller scale building a binary tree structure
as shown in figure \ref{fig:branch}.  In this way the number of structures increases exponentially as smaller scales are reached. Such models with binary structure were introduced in the 1990s but have not been investigated
extensively \cite{aurell1994hierarchical,aurell1997binary}. Here, we follow a similar analysis as in \cite{biferale1994chaotic} where stationary solutions of non-binary shell models were investigated. We demonstrate that such analysis allows the construction of exact stationary solutions that display intermittency.

%%%%%%%%%%%%%%%%%%%%%%%%%%%%%%%%%%%%%%%%%%%%%%%%%%%%%%%%%%%%%%%%%%%%%%%%%%%%%%%%%%%%%%%%%%%%%%%%%%%%%%%%%%%%%
\section{ Multi-branch shell models }   %%%%%%%%%%%%%%%%%%%%%%%%%%%%%%%%%%%%%%%%%%%%%%%%%%%%%%%%%%%%%%%%%%%%%
%%%%%%%%%%%%%%%%%%%%%%%%%%%%%%%%%%%%%%%%%%%%%%%%%%%%%%%%%%%%%%%%%%%%%%%%%%%%%%%%%%%%%%%%%%%%%%%%%%%%%%%%%%%%%

%%%%%%%%%%%%%%%%%%%%%%%%%%%%%%%%%%%%%%%%%%%%%%%%%%%%%%%%%%%%%%%%%%%%%%%%%%%%%%%%%%%%%%%%%%%%%%%%%%%%%%%%%%%%%
\begin{figure}[H]                                                                                      %%%%%%
%\begin{adjustwidth}{-\extralength}{0cm}
\centering
\includegraphics[width=12.5cm]{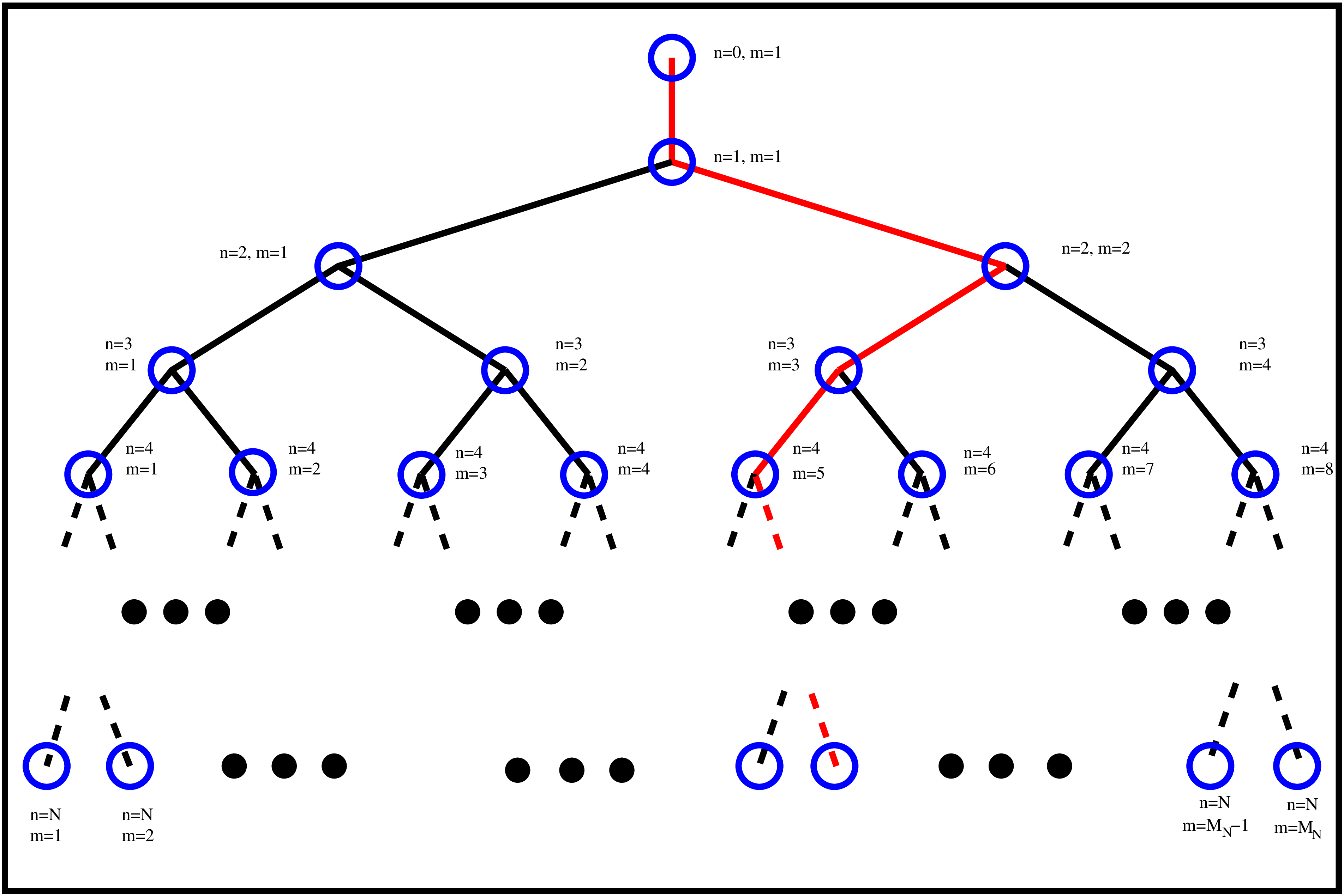}
%\end{adjustwidth}
\caption{A sketch of the two branch ($\mu=2$) shell model. Each node marked by blue circles represents one dynamical mode of amplitude $U_{n,m}$ marked by the two indexes $n,m$ where $n$ characterizes the scale $\ell_n = \ell_0 \lambda^n$ and $m$ characterises the number of the mode in that scale. In each scale $n$ there are $M_n=\mu^{n-1}$ modes. 
%The index $i$ counts the number of modes at all scales it goes from $1$ to $(\mu^N-1)/(\mu-1)$ and it is used in the numerical procedure to solve the shell model described in the appendix. 
\label{fig:branch}}
\end{figure} 
%\unskip                                                                                                %%%%%%
%%%%%%%%%%%%%%%%%%%%%%%%%%%%%%%%%%%%%%%%%%%%%%%%%%%%%%%%%%%%%%%%%%%%%%%%%%%%%%%%%%%%%%%%%%%%%%%%%%%%%%%%%%%%%

%\section{Basic Construction}

We consider the evolution of a turbulent flow modeled by the 
%(real or complex) 
real amplitudes 
$U_{n,m}$ of structures of scale $\ell_n=1/k_n$ where 
%%%
\begin{equation}
k_n=\lambda^n k_0 \quad \mathrm{or} \quad \ell_n= \ell_0/\lambda^n 
\end{equation}
%%%
and $1<\lambda$.
At scale $\ell_1$ there is one structure whose amplitude is given by $U_{1,1}$, this structure will transfer its energy
to $\mu \in \mathbb{N}$ structures of scale $\ell_2$, each one of which will transfer its energy to $\mu$ structures of scale 
$\ell_3$ and so on as shown in figure \ref{fig:branch} for $\mu=2$. The volume of each structure is given by $V_i=\ell_n^D$ where $D$ is the spatial dimension.
If the cascade process is space filling the number of substructures $\mu$ is related to $\lambda$ and $D$ by 
%%%
\begin{equation} 
\lambda^D=\mu.
\end{equation} 
%%%
Accordingly at energy scale $\ell_i$ we have $M_n=\mu^{n-1}$ (with $M_0=1$) structures so that if we consider 
$N$ such scales we have a total of
%%%
\begin{equation}
Z = 1+ \sum_{n=1}^{N} M_n = \frac{\mu^{N}-1}{\mu-1}+1 
\end{equation}
%%%
structures.
%\[ k_i = k_0 \lambda^i \]
The energy of every structure is given by 
%%%
\begin{equation} 
E_{n,m}=\frac{1}{2}\rho U^2_{n,m}V_n = \frac{1}{2}\rho U^2_{n,m} \ell_n^D 
\end{equation}
%%%
so the total energy is given by 
\begin{equation}
E=\frac{1}{2}\sum_{n=0}^N \frac{1}{M_n}\sum_{m=1}^{M_n}  U^2_{n,m}
\label{eq:ener}
\end{equation}
where $\rho$ is from now on taken to be unity.
%where $V_i=\ell_i^D$ is the $D$-dimensional volume occupied by that structure. 
%This energy cascades to $\mu \in \mathbb{N}$ substructures.

%\section{Model equations}
%\subsection{Nearest neighbours one branch model  model (Desnianskii \& Novikov model) }
%
In the  Desnianskii \& Novikov model \cite{desnianskii1974simulation} structures of scale $\ell_i$ interact with only
structures of scale $\ell_{n+1}$ and $\ell_{n-1}$ and there is no branching $\mu=1$. 
%
%\subsection{Nearest neighbours one branch model $\mu=1$}
%
%Considering the simplest possible case of  one branch and only nearest neighbour's interactions
The amplitudes $U_{n}$ then follow the following dynamical equation:
\begin{equation}
\Dot{U}_{n} = a k_n[ U_{n-1}U_{n-1} - \lambda U_nU_{n+1} ] +
              b k_i[ U_{n}U_{n-1} - \lambda U_{n+1}U_{n+1} ] - \nu k^2U_n + F_n
\end{equation}
For $\nu=0$ and $F_n=0$ this system conserves the energy \ref{eq:ener} (with $M_n=1$)
for any value of $a,b$. %:
The flux of energy across a scale $\ell_n$ is given by:
\begin{eqnarray}
\Pi_{n}  &=&   a k_n U_{n} U_{n-1} U_{n-1} + b k_n  U_{n,m}U_{n,m} U_{n-1} .
\end{eqnarray}
%\begin{equation}
%E = \frac{1}{2}  \sum_{n=1}^N U_n^2.
%\end{equation}
%%%%%%%%%%%%%%%%%%%%%%%%%%%%%%%%%%%%%%
%\subsection{Nearest neighbours two-branch model $\mu=2$}
%%%%%%%%%%%%%%%%%%%%%%%%%%%%%%%%%%%%%%

Expanding on the Desnianskii \& Novikov model allowing each structure to branch out to two ($\mu=2$) smaller scale structures
%Considering the simplest possible case and only nearest neighbour's interactions
$U_{i,j}$ results in the following dynamical equation:
\begin{eqnarray}
    \Dot{U}_{n,m} &=& 
     a {k_n} \left[ U_{n-1,m^*} U_{n-1,m^*} - 
     \frac{\lambda}{2}  \left(
        U_{n,m}U_{n+1,m'}+U_{n,m}U_{n+1,m'+1} 
       \right) \right] + \\
                  & & 
     b {k_n} \left[ U_{n,m} U_{n-1,m^*} -  
     \frac{\lambda}{2} \left(
      U_{n+1,m'}U_{n+1,m'}+U_{n+1,m'+1}U_{n+1,m'+1}
      \right) \right] \\ & & -\nu k_n^2 U_{n,m}+ F_{n,m}
      \label{eq:model}
\end{eqnarray}
where   $\nu$ is the viscosity, $F_{n,m}$ is the forcing and $a,b$ are again free parameters.
The branching diagram for the model given in \ref{eq:model} is given in figure \ref{fig:branch}. 
The integer $m'$ and $m'+1$ correspond to the index of scales $\ell_{n+1}$ with which the mode $U_{n,m}$ is linked 
where $m'$ is explicitly given my $m'=2m-1$ and $m^*$ corresponds to the index of scale $\ell_{n-1}$ linked to $U_{n,m}$
given by $m^*=\mathrm{Int}[(m+1)/2]$ as illustrated in the left panel of figure \ref{fig:circle}.
%
%\noindent {\bf Proposition:} 
For $\nu=0$, $F_{n,m}=0$ and for any value of $a,b$ the system conserves the energy \ref{eq:ener}
where now $M_n=2^{n-1}$.
%\begin{equation}
%E = \sum_{n=0}^{N} \sum_{m=1}^{M_n} E_{n,m}=\frac{1}{2}\rho  \sum_{n=0}^{n=N} 
%\left( \frac{1}{M_n} \sum_{m=1}^{M_n} U^2_{m,n} \right).
%\end{equation}
%
%\noindent
%{\bf Energy flux:}
The energy flux $\Pi_{n,m}$ through a scale $\ell_n$ and structure ($n,m$) (expressing the rate energy from the large scales ($i < n$) is lost to the smaller scales
($i \ge n$) through the structure $m$ due to the non-linearity is given by 
\begin{eqnarray}
\Pi_{n,m}  &=& %\sum_{i=n}^N  \sum_{j=...}^{...} U_{i,j} \Dot{U}_{i,j}   \\
       %&=& \sum_{i=n}^N   \dots \\
      % &=& 
              a k_n U_{n,m} U_{n-1,m^*} U_{n-1,m^*} + b k_n  U_{n,m}U_{n,m} U_{n-1,m^*}  \label{eq:flux}
\end{eqnarray}
The total flux through scale $\ell_n$ is then given by
\begin{equation} 
\Pi_n = \frac{1}{M_n}\sum_{m=1}^{M_n} \Pi_{n,m} \label{eq:fluxT}
\end{equation}
Conservation of energy by the non-linear terms implies that at scales smaller than the  forcing scale 
and larger than the dissipation scale ($\ell_\nu$) the flux $\Pi_n$ is constant and equal to
the energy injection/dissipation $\epsilon$
\begin{equation} 
\Pi_n = \epsilon , \qquad 1<n\ll n_\nu
\end{equation}
where $\ell_\nu = (\nu^3/\epsilon)^{1/4} $ and $n_\nu = \log_\lambda(\ell_1/\ell_\nu).$
The range $1<n\ll n_\nu$ where forcing and viscous effects can be neglected is called the inertial range.

%%%%%%%%%%%%%%%%%%%%%%%%%%%%%%%%%%%%%%%%%%%%%%%%%%%%%%%%%%%%%%%%%%%%%
\begin{figure}[H]                                               %%%%%
\begin{center}
\includegraphics[width=6.5 cm]{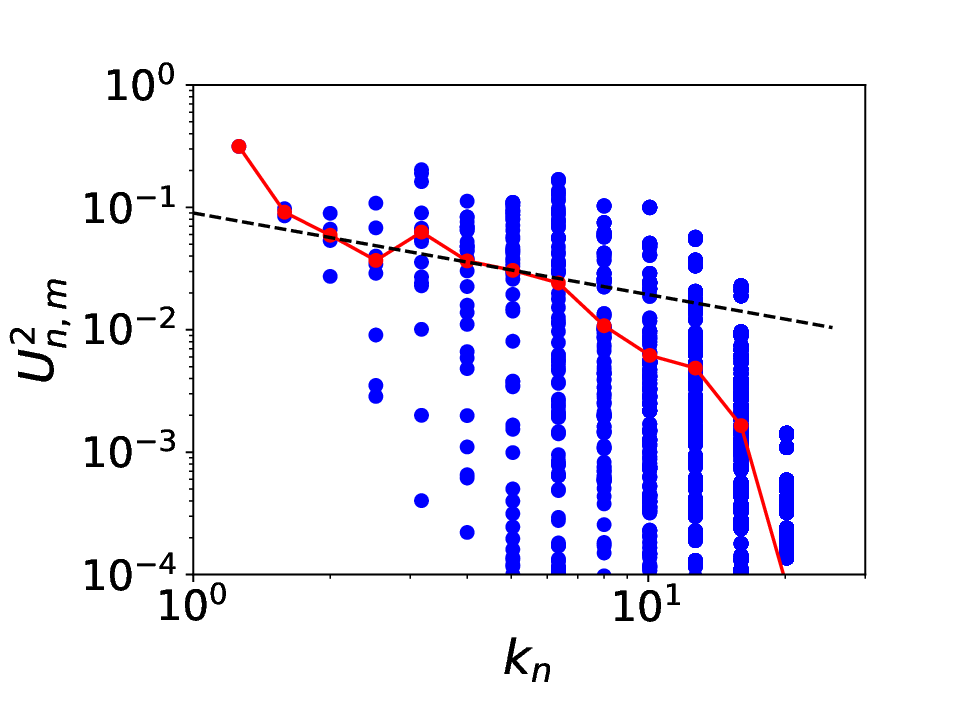}
\includegraphics[width=6.5 cm]{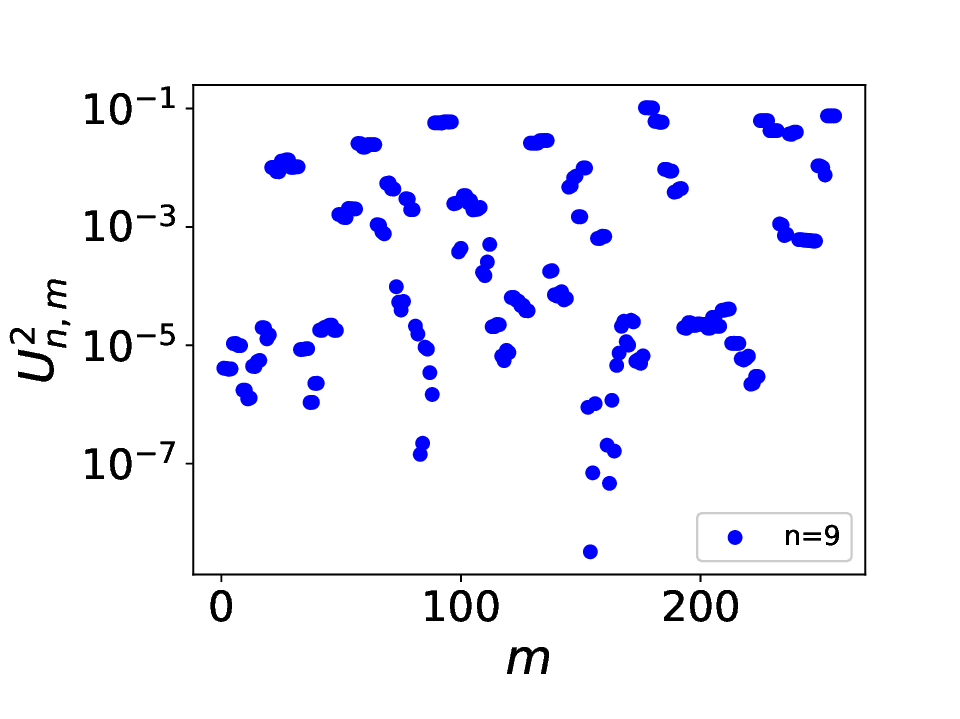}
\caption{ Energy spectrum from a numerical simulations of the model \ref{eq:model}. In the left panel the red points correspond to $U_{n,m}$ averaged over $m$ for a given $n$ 
while the blue points correspond $U_{n,m}$ for all $n,m$.  The right panel displays $U_{n,m}$ as a function of $m$ for $n=9$. \label{fig:Realisation} \\ }
\end{center}
\end{figure}   
%\unskip                                                        %%%%%
%%%%%%%%%%%%%%%%%%%%%%%%%%%%%%%%%%%%%%%%%%%%%%%%%%%%%%%%%%%%%%%%%%%%%
In figure \ref{fig:Realisation} we plot the energy spectra $U^2_{n,m}$ as a function of 
$n$ with blue dots, while the red dots indicate the averaged value $\overline{U^2_{n}}=(\sum_m U^2_{n,m})/M_n$
from a realisation of a simulation of eq. \ref{eq:model} performed with $N=14, \lambda=2^{1/3}$ forced at $n=1$.
The averaged value follows power-law close to the Kolmogorov scaling $\overline{U^2_{n}}\propto k_n^{-2/3}$ although
individual $U^2_{n,m}$ can vary orders of magnitude from this mean value. This indicates that higher order statistics
can deviate from the dimensional analysis spectrum. The present model is computationally expensive as its complexity
increases as $2^N$. As a result it is not easy to obtain a long inertial range (large $N$) to investigate 
the resulting power-law behaviors numerically. On the other hand however its simplicity allows for analytical 
treatment which is what we are examining in the next section by constructing exact inertial range solutions 
of arbitrary large $n$.

%%%%%%%%%%%%%%%%%%%%%%%%%%%%%%%%%%%%%%%%%%%%%%%%%%%%%%%%%%%%%%%%%%%%%%%%%%%%%%%%%%%%%%%%%%%%%%%%%%%%%%%%%%%%%%%
%%%%%%%%%%%%%%%%%%%%%%%%%%%%%%%%%%%%%%%%%%%%%%%%%%%%%%%%%%%%%%%%%%%%%%%%%%%%%%%%%%%%%%%%%%%%%%%%%%%%%%%%%%%%%%%
\section{Inertial range intermittent solutions}   %%%%%%%%%%%%%%%%%%%%%%%%%%%%%%%%%%%%%%%%%%%%%%%%%%%%%%%%%%%%%
%%%%%%%%%%%%%%%%%%%%%%%%%%%%%%%%%%%%%%%%%%%%%%%%%%%%%%%%%%%%%%%%%%%%%%%%%%%%%%%%%%%%%%%%%%%%%%%%%%%%%%%%%%%%%%%
%%%%%%%%%%%%%%%%%%%%%%%%%%%%%%%%%%%%%%%%%%%%%%%%%%%%%%%%%%%%%%%%%%%%%%%%%%%%%%%%%%%%%%%%%%%%%%%%%%%%%%%%%%%%%%%

%%%%%%%%%%%%%%%%%%%%%%%%%%%%%%%%%%%%%%%%%%%%%%%%%%%%%%%%%%%%%%%%%%%%%
\begin{figure}[H]                                               %%%%%
\includegraphics[width=6.5 cm]{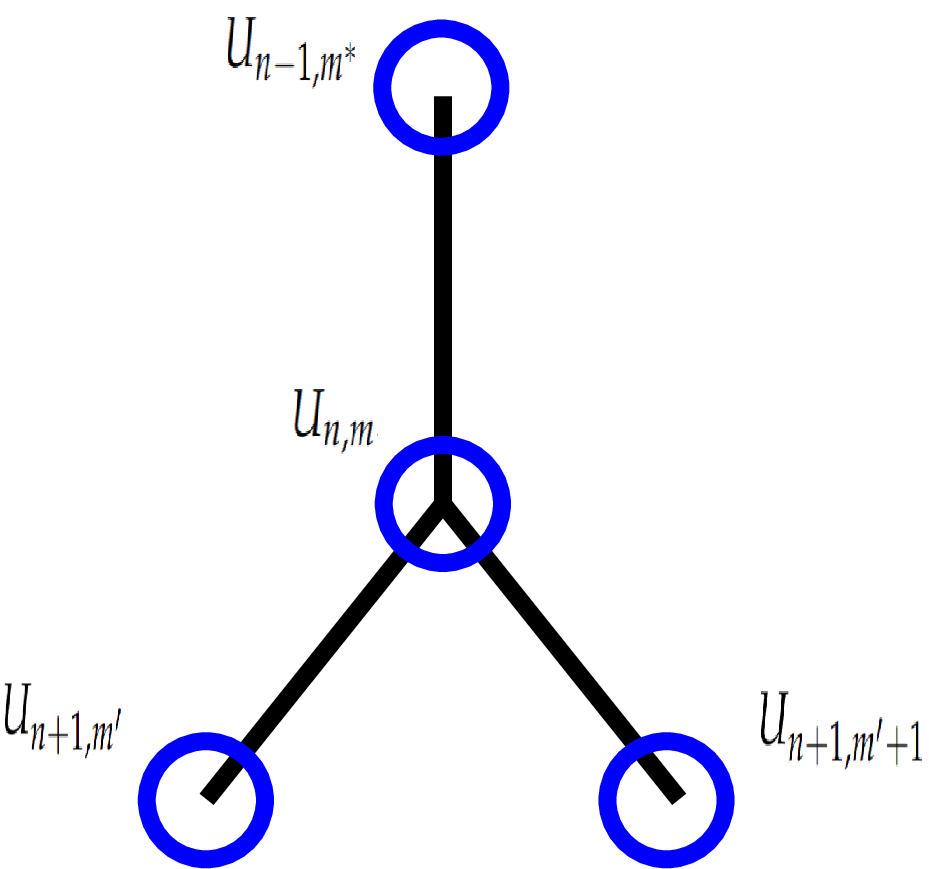}
\includegraphics[width=6.5 cm]{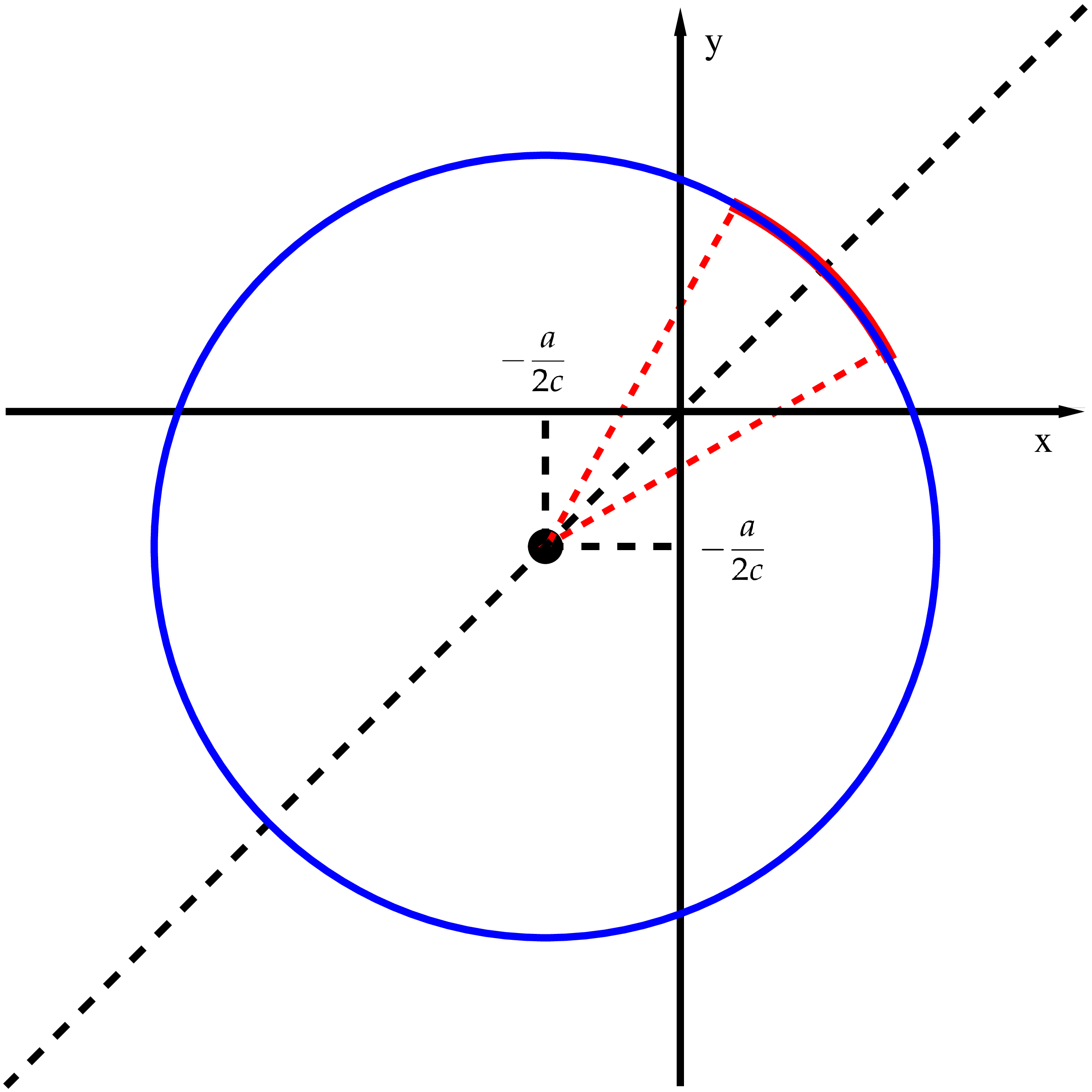}
\caption{This is a wide figure.\label{fig:circle} \\ }
\end{figure}   
\unskip                                                         %%%%%
%%%%%%%%%%%%%%%%%%%%%%%%%%%%%%%%%%%%%%%%%%%%%%%%%%%%%%%%%%%%%%%%%%%%%

We look for stationary solutions of eq. \ref{eq:model} such that in the inertial range where forcing and
dissipation can be ignored. Stationarity  implies that for any $n,m$:
\begin{eqnarray}
0 &=& a \left[ U_{n-1,m^*} U_{n-1,m^*} - 
      \frac{\lambda}{2}  \left(
        U_{n,m}U_{n+1,m'}+U_{n,n}U_{n+1,m'+1} 
       \right) \right] \nonumber \\ 
       &+&            
     b \left[ U_{n,m} U_{n-1,m^*} -  
     \frac{\lambda}{2} \left(
      U_{n+1,m'}U_{n+1,m'}+U_{n+1,m'+1}U_{n+1,m'+1}
      \right) \right]
      \label{eq:stationary1}
\end{eqnarray}
The way we proceed to find such a solution is the following:
Given $U_{n-1,m^*}$ and $U_{n,m}$ we look for $U_{n+1,m'}$ and $U_{n+1,m'+1}$ such that the
equation above is satisfied; then we proceed to the next scale and search for  $U_{n+2,2m'-1}$ and $U_{n+2,2m'}$
and so on finding a recursive relation that gives all $U_{n,m}$.
%Let %
The solutions only depend on the relative amplitude of $U_{n,m}$ so 
we define their normalised ratio as:
\begin{equation}
%r=
r_{n,m} = \frac{U_{n,m}\lambda^{1/3}}{U_{n-1,m^*}} 
%, \quad x =r_{n+1,m'} = \frac{U_{n+1,m'}\lambda^{1/3}}{U_{n,m}}, \quad y =r_{n+1,m'+1} =\frac{U_{n+1,m'+1}\lambda^{1/3}}{U_{n,m}}
%, \quad \mathrm{and} \quad b=c\lambda^{1/3}.
\end{equation}
To simplify the notation we denote 
\begin{equation}
r=r_{n,m}, \quad x=r_{n+1,m'} \quad y=r_{n+1,m'+1} \quad \mathrm{and} \quad b=c\lambda^{1/3}
\label{eq:ratio}
\end{equation}
then stationary solutions of \ref{eq:stationary1} satisfy
\begin{eqnarray}
0 &=& U_{n,m}^2 \lambda^{2/3} \left( a \left[ \frac{1}{r^2} - \frac{1}{2}\left( x  +y  \right)   \right]  + 
                                     c \left[ \frac{1}{r  } - \frac{1}{2}\left( x^2+y^2\right)   \right] \right)
\end{eqnarray}
that simplifies to
\begin{eqnarray}
 \left( x+ \frac{a}{2c}\right)^2+ \left( y+\frac{a}{2c} \right)^2 = 2 \left( \frac{a}{cr^2}+  \frac{1}{r} + \frac{a^2}{4c^2}  \right). \label{eq:xy}
\end{eqnarray}
which has real solutions only if
\begin{eqnarray}
 0 \le  \frac{a}{cr^2}+  \frac{1}{r} + \frac{a^2}{4c^2} =R^2  \label{eq:c1}.
\end{eqnarray}
The solutions $x,y$ form  a circle in the $x,y$ plane centered at $-a/2c,a/2c$ and radius $R$
depicted in the right panel of figure \ref{fig:circle}. It is important to note that any point $x,y$ 
in this circle is a solution of \ref{eq:xy}, and thus we have multiple possible solutions.
The condition \ref{eq:c1} is satisfied for positive $r,a,c$
that will be the focus of the present investigation. Returning to the $r_{n,m}$ notation 
the values of $r_{n+1,m'}$ and $r_{n+1,m'+1}$ that satisfy the stationarity condition can be written 
in full generality as:
\begin{eqnarray}
r_{n+1,m'} &=&-\frac{a}{2c}+\sqrt{2}\cos(\theta_{n,m}) \sqrt{ \frac{a}{cr_{n,m}^2}+  \frac{1}{r_{n,m}} + \frac{a^2}{4c^2}}  \label{eq:solx}\\ 
r_{n+1,m'+1} &=&-\frac{a}{2c}+\sqrt{2}\sin(\theta_{n,m}) \sqrt{ \frac{a}{cr_{n,m}^2}+  \frac{1}{r_{n,m}} + \frac{a^2}{4c^2}} \label{eq:soly}
\end{eqnarray}
where $\theta_{n,m}$ is arbitrary. Equations \ref{eq:solx},\ref{eq:soly} form a recurrence relation out of which given $r_{1,1}$ and a choice of $\theta_{n,m}$ one can construct all $r_{n,m}$.  Then, given $r_{n,m}$ one can obtain $U_{n,m}$ based on \ref{eq:ratio} as 
\begin{equation} 
U_{n,m} = U_{1,1}\, r_{1,1}\,r_{2,m_1}\, r_{3,m_2} \, \dots \, r_{n,m} \label{eq:UR}
\end{equation}
where $m_1,m_2,\dots$ are the $m$ one crosses along the path from (1,1) to ($n,m$) as shown by the red line in figure \ref{fig:branch}.  
This recurrence relation however does not always lead to bounded solutions of $r_{n,m}$.
For some values of $\theta_{n,m}$ the resulting $x,y$ can be negative or zero. Negative values can
lead to un-physical solutions with negative flux from the small to the large scales which
are not possible (for stationary solutions) since no energy source is assumed at small scales.
If $x$ or $y$ is zero it means that the particular branch is zero for all subsequent values. 
We need thus to limit the choice of $\theta$ so that positive and finite $r_{n,m}$ are obtained. 

The simplest case is obtained by choosing $\theta_{n,m}=\pi/4$. It corresponds to an equal part of energy
being distributed to the left and the right branch and leads to the Kolmogorov solution $r_{n,m}=1$
or in terms of the velocity $U_{n,m}=\lambda^{n/3}$) (where $U_{1,1}=1$ is assumed). It corresponds to a finite flux
non-intermittent (self-similar) solution. 
%It can be shown that if $r_{1,1}\ne1$ is chosen $r_{n,m}$ converges to $r_{n,m}\to1$ if $c<a$, to to $0,\infty$ in a two step loop if $a>c$ and oscillates around $1$ if $a=c$. For the $c<a$ case the only acceptable solution is the $r_{n,m}=1$. In all cases asymptotically the solutions follow the Kolmogrov scaling and display no intermittency.

Intermittency however can manifest itself if we chose $\theta_{n,m}\ne \pi/4$ so that energy is not equally 
distributed in the left and right branch. 
Here we will chose $\theta_{n,m}$  randomly with uniform distribution in the range
$\theta_{\min}=\pi/4-\Delta \theta < \theta_{n,m} <\pi/4+\Delta \theta=\theta_{\max}$ for a given $\Delta \theta<\pi/4$.
Then it can be shown that for $c>a$ there exists $r_{\max}>r_{\min}>0$ such that for all $r\in (r_{\max},r_{\min})$ both
$x\in (r_{\max},r_{\min}) $ and $y\in (r_{\max},r_{\min}) $. For $c\le a$ the recurrence relation converges either to
$r_{n,m}=0$ or $r_{n,m}=\infty$ and we are going to limit ourselves only to the $c>a$ case here.
To obtain $r_{\max},r_{\min}$ one needs to note that from the recurrence relation \ref{eq:soly}
the largest value of $r_{n+1,m'}=r_{max}$ is obtained when $\theta=\theta_{max}$ and $r_{n,m}=r_{min}$
while the smallest value of $r_{n+1,m'}=r_{min}$ is obtained when $\theta=\theta_{min}$ and $r_{n,m}=r_{max}$.
This leads to the following relations
\begin{eqnarray}
r_{\max} &=&-\frac{a}{2c}+\sqrt{2}\cos(\theta_{\min})\sqrt{ \frac{a}{cr_{\min}^2}+  \frac{1}{r_{\min}} + \frac{a^2}{4c^2}}
\label{eq:rmax}\\ 
r_{\min} &=&-\frac{a}{2c}+\sqrt{2}\cos(\theta_{\max})\sqrt{ \frac{a}{cr_{\max}^2}+  \frac{1}{r_{\max}} + \frac{a^2}{4c^2}} 
\label{eq:rmin}.\\
\end{eqnarray}
We arrive at exactly the same relations if we examine eq. \ref{eq:soly}.
%(taking into account that $\cos(\theta_{\min})=\sin(\theta_{\max})$ and $\cos(\theta_{\max})=\sin(\theta_{\min})$).
%%%%%%%%%%%%%%%%%%%%%%%%%%%%%%%%%%%%%%%%%%%%%%%%%%%%%%%%%%%%%%%%%%%%%%%%%%%
\begin{figure}[H]                                                     %%%%%
\includegraphics[width=0.45 \textwidth]{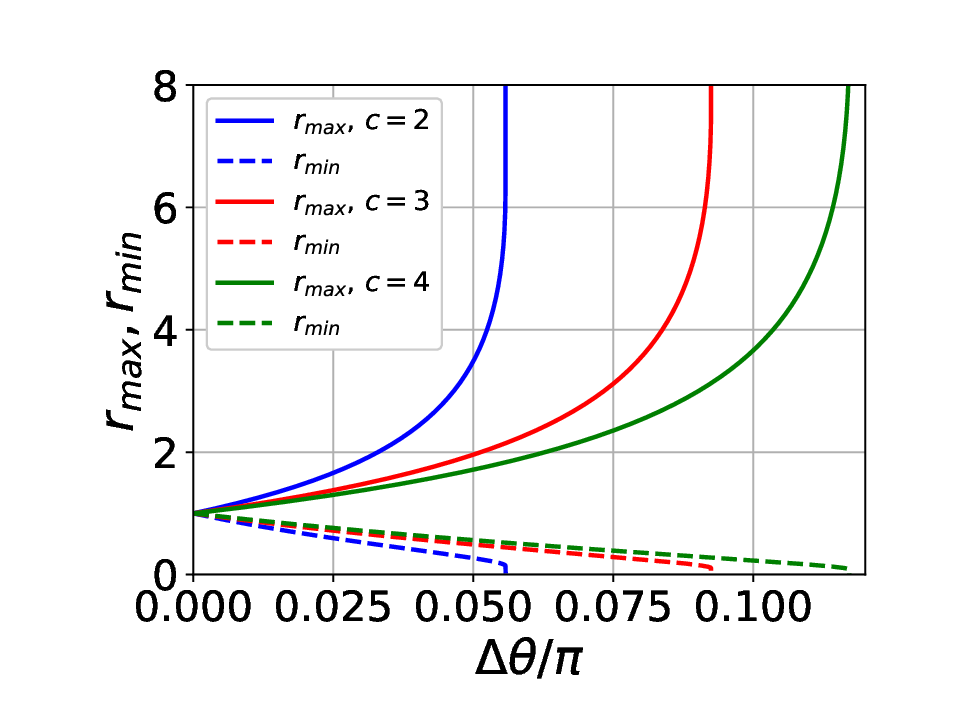}
\includegraphics[width=0.45 \textwidth]{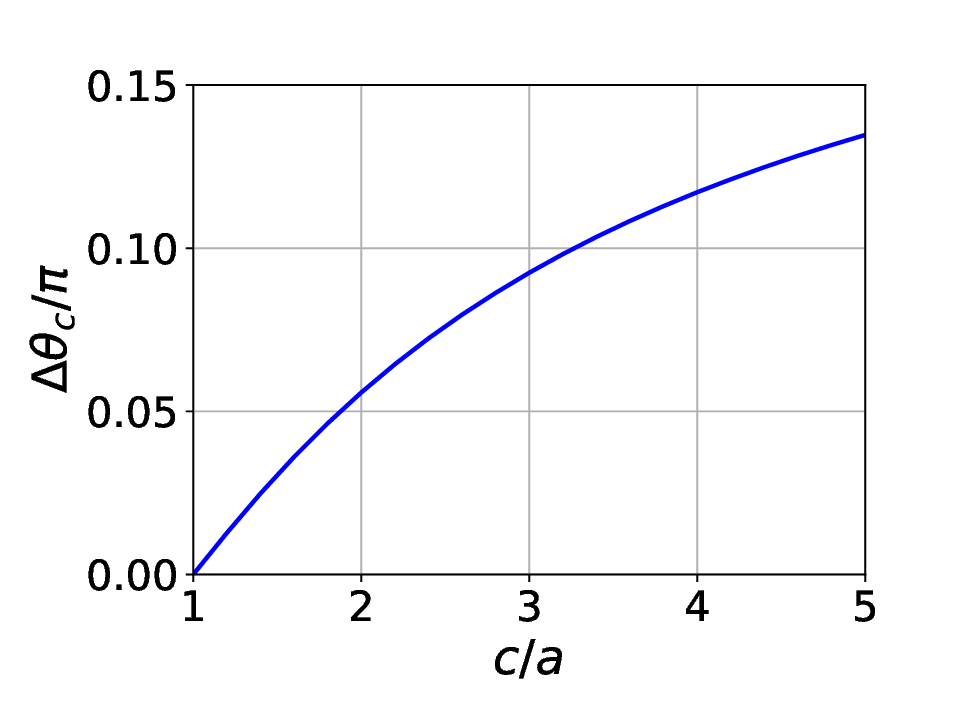}
\caption{....... \label{fig:theta} \\ }
\end{figure}   
\unskip                                                                %%%%%
%%%%%%%%%%%%%%%%%%%%%%%%%%%%%%%%%%%%%%%%%%%%%%%%%%%%%%%%%%%%%%%%%%%%%%%%%%%%

We solved equations \ref{eq:rmax},\ref{eq:rmin} numerically and the results are shown in the left panel of  figure \ref{fig:theta} for three different values of $c/a$. For $\Delta \theta =0 $ only the Kolmogorov solution is allowed 
with $r_{max}=r_{min}=1$. As $\Delta \theta =0 $ increases $r_{n,m}$ cover a wider range of values 
up until a critical value of $\Delta \theta =\Delta \theta_c $ for which $r_{min}$ becomes zero
and $r_{max}$  diverges. The value of this critical angle as a function $c/a$ is shown in the right panel
of the same figure. $\Delta \theta_c$ is zero for $c/a=1$ and grows for larger values approaching
$\Delta \theta_c=\pi/4$ as $c/a\to \infty$ (not demonstrated here). 

For any given choice of $\Delta \theta < \Delta \theta_c$ we can construct an ensemble of exact solutions of the
present model by following the recurrence relations \ref{eq:solx},\ref{eq:soly} picking each time randomnly $\theta_{n,m}\in (\pi/4-\Delta \theta, \pi/4+\Delta \theta )$ and reconstructing $U_{n,m}$ by eq. \ref{eq:UR}. We note that other than $c/a$ the only 
other parameter that controls the ensemble of solutions considered is $\Delta \theta/\Delta \theta_c$ that 
provides a measure of how much our ensemble deviates from the Kolmogorov solution $\Delta \theta=0$.
This process has direct links with the random cascade models studied in the past \cite{novikov1964turbulent,yaglom1966effect,mandelbrot1999intermittent}, however we need to note that
unlike the random cascade models the solutions found here are energy conserving.

%%%%%%%%%%%%%%%%%%%%%%%%%%%%%%%%%%%%%%%%%%%%%%%%%%%%%%%%%%%%%%%%%%%%%%%%%%%%%%%%%%%%%%%%%%%%%%%%%%%%%%%%%%%%
%%%%%%%%%%%%%%%%%%%%%%%%%%%%%%%%%%%%%%%%%%%%%%%%%%%%%%%%%%%%%%%%%%%%%%%%%%%%%%%%%%%%%%%%%%%%%%%%%%%%%%%%%%%%
\section{ Statistical behavior and Intermittency}  %%%%%%%%%%%%%%%%%%%%%%%%%%%%%%%%%%%%%%%%%%%%%%%%%%%%%%%%%
%%%%%%%%%%%%%%%%%%%%%%%%%%%%%%%%%%%%%%%%%%%%%%%%%%%%%%%%%%%%%%%%%%%%%%%%%%%%%%%%%%%%%%%%%%%%%%%%%%%%%%%%%%%%
%%%%%%%%%%%%%%%%%%%%%%%%%%%%%%%%%%%%%%%%%%%%%%%%%%%%%%%%%%%%%%%%%%%%%%%%%%%%%%%%%%%%%%%%%%%%%%%%%%%%%%%%%%%%

In this section we examine a large ensemble of the exact solutions shown in the previous section and investigate their properties.
For our investigation we have set $c/a=2$ and we consider only a single path (as the one shown in red in figure \ref{fig:branch}) and not the full tree. The differences in the statistics between the two choices (single path and full tree) 
lie in the cross correlations between different modes that are not captured in the single path. As an example 
we mention that the flux  $\Pi_n$ in eq. \ref{eq:fluxT} is identically equal to $\epsilon$ for every realization 
while the flux $\Pi_{n,m}$ given in eq.\ref{eq:flux} fluctuates and only its mean value is equal to $\epsilon$
\[ \langle \Pi_{n,m}\rangle = \Pi_n=\epsilon.\]
Along such path we consider three different ensembles for $\Delta \theta / \Delta \theta_c = 0.1, \, 0.5, \, 0.9 $ each one composed of 
$10^7$ different solutions. The solutions were constructed by picking randomly  $\theta_{n,m}$ for each node examined, 
from a uniform distribution between $\pi/4- \Delta\theta $ and $\pi/4 + \Delta\theta $. The value of $n$ 
varied from $n=1$ to $n=200$. We note that if the full tree was investigated instead of a single path 
for such large value $n$ it would require to solve for $2^{200}$ degrees of freedom that is computationally unattainable.

\begin{figure}[H]
\centering
\includegraphics[width=0.3 \textwidth]{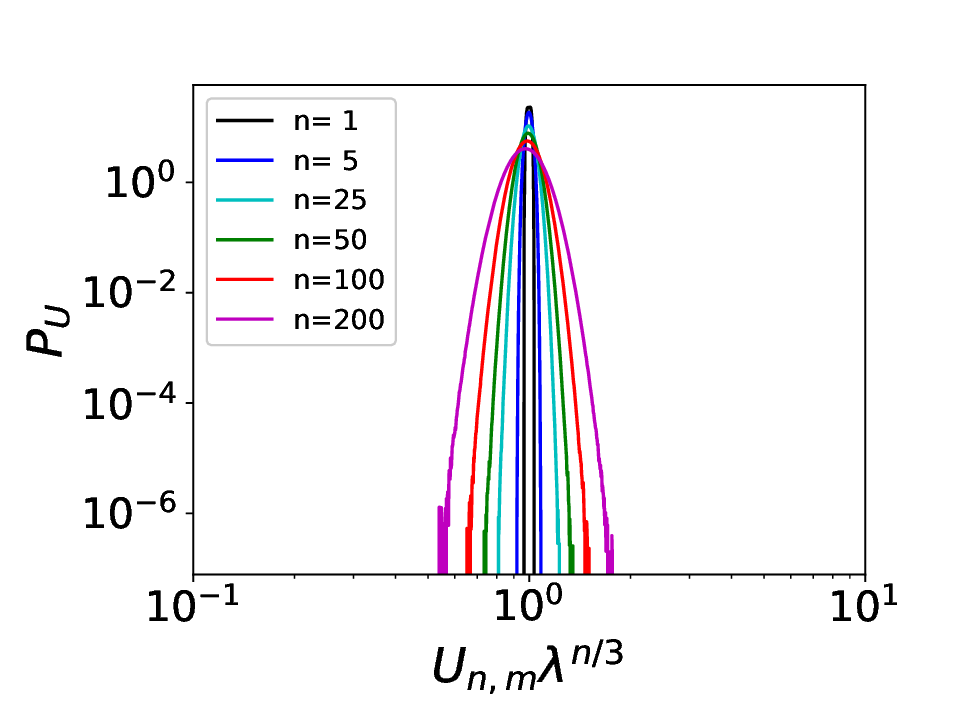}
\includegraphics[width=0.3 \textwidth]{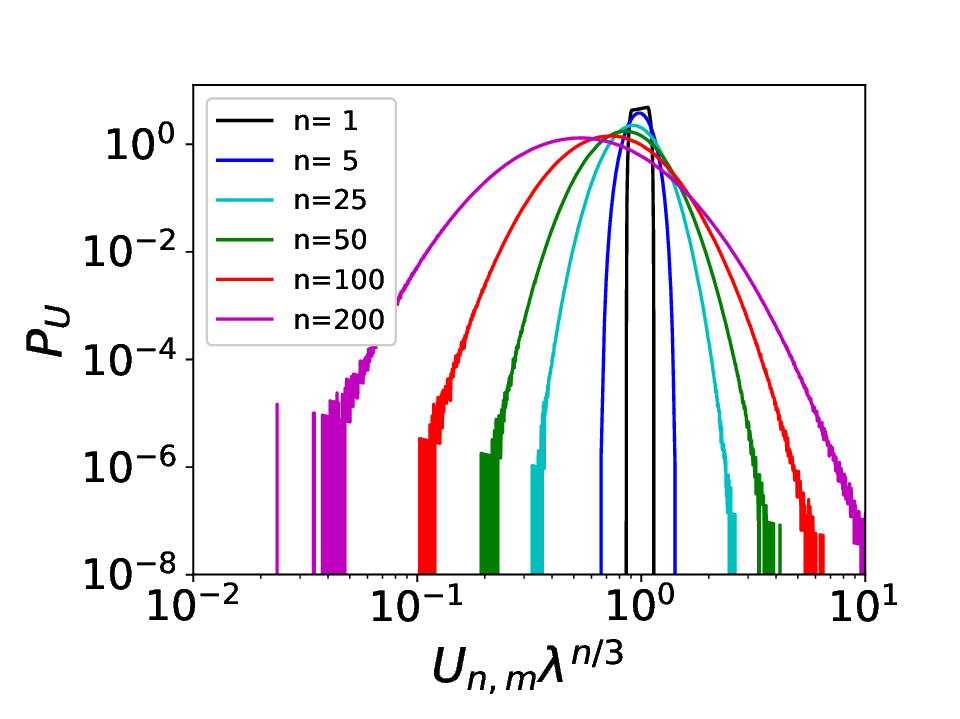}
\includegraphics[width=0.3 \textwidth]{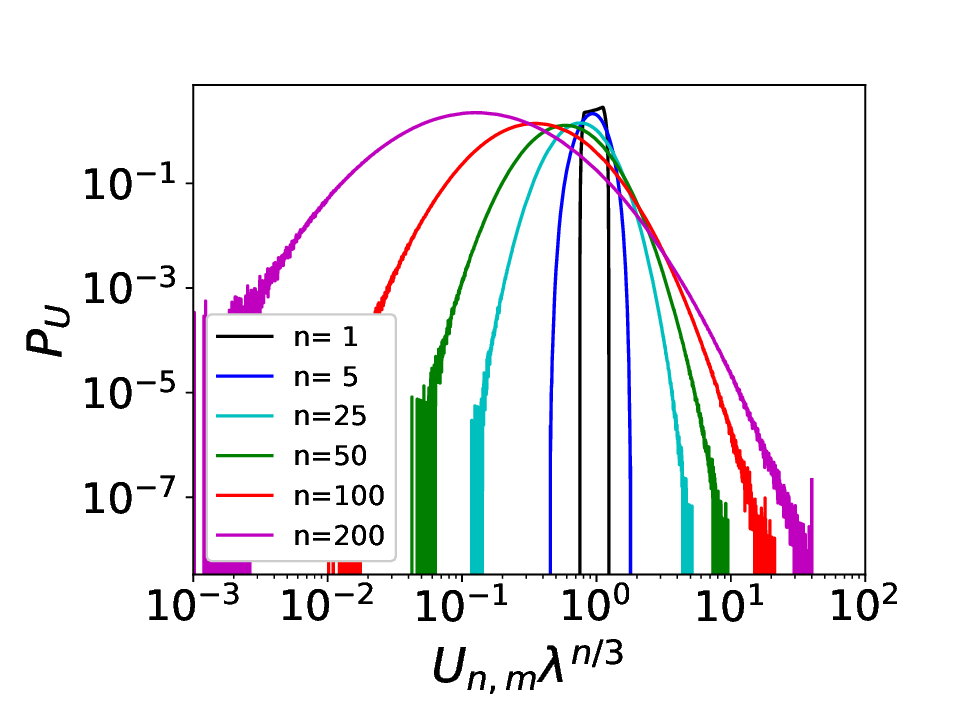}
\centering
\includegraphics[width=0.3 \textwidth]{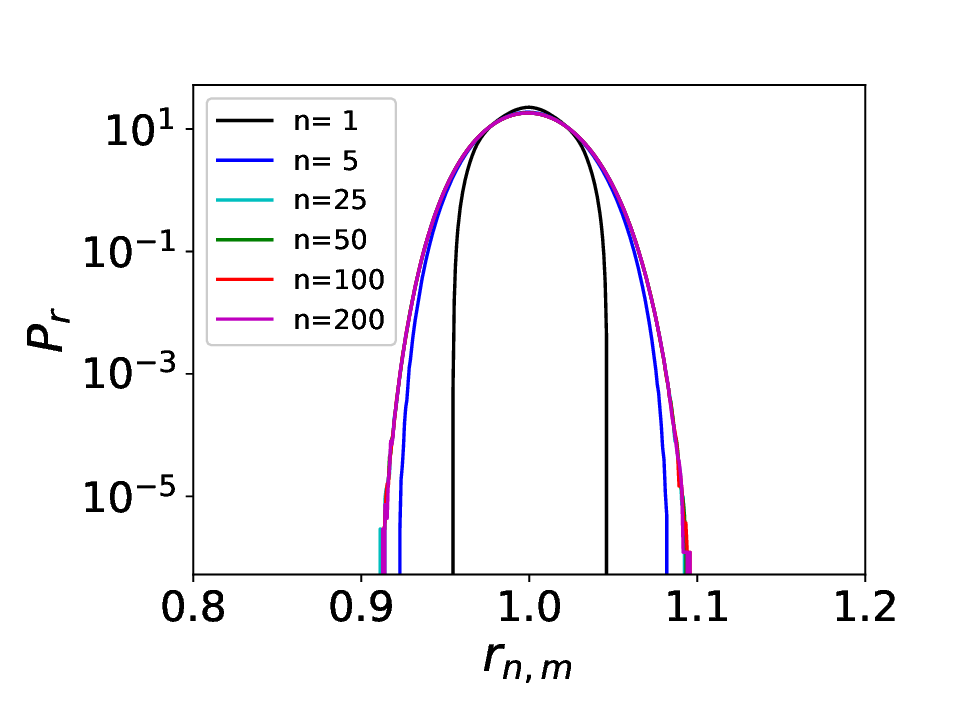}
\includegraphics[width=0.3 \textwidth]{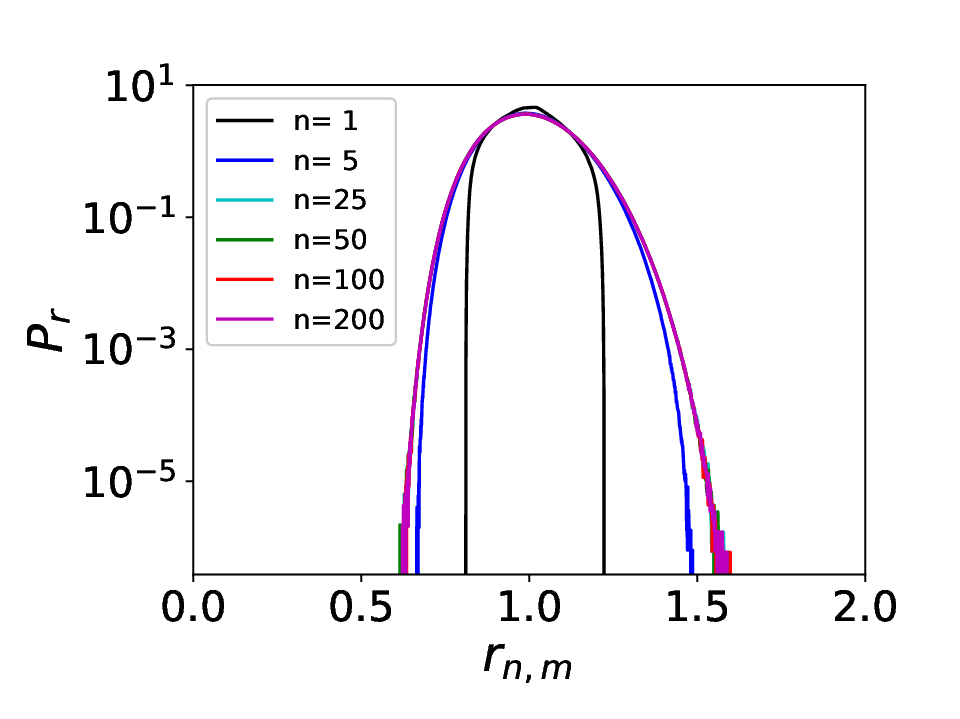}
\includegraphics[width=0.3 \textwidth]{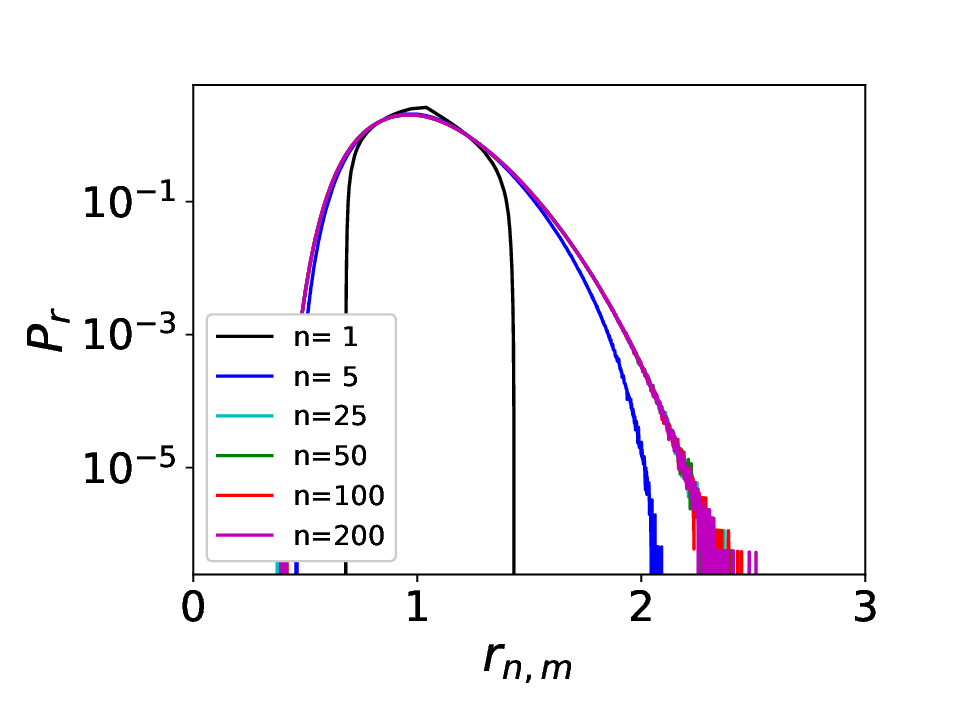}
\caption{{\bf Top panels:} PDFs $P_U(U_{n,m})$ of the velocity modes $U_{n,m}$ for the three different ensembles (left $\Delta \theta / \Delta \theta_c = 0.1$, center $\Delta \theta / \Delta \theta_c = 0.5 $ and right $\Delta \theta / \Delta \theta_c = 0.9$) for different values of $n$.
{\bf Bottom panels:} PDFs $P_r(r_{n,m})$ of the velocity ratios $r_{n,m}$ for the same ensembles and the same $n$.
\label{fig:PDFUU}}
\end{figure}   
%\unskip
In the top panels of figure \ref{fig:PDFUU} we plot the probability distribution function (PDF) $P_U(U_{n,m})$ of the variable
$U_{n,m}$ for the three values of $\Delta \theta / \Delta \theta_c = 0.1, \, 0.5, \, 0.9$ (from left to right) and different values of $n$. 
The PDFs of different values of $n$ do not seem to overlap, although the $x$-axis has been normalised by the Kolmogorov prediction $\lambda^{-1/3}$. 
Instead as large values of $n$ are reached the pdf's display larger tails reaching values of $U_{n,m}$ much larger and much smaller than its mean value. 
The closer $\Delta \theta$ is to the critical value $\Delta \theta_c$ the larger this deviation is.
On the other hand, the PDFs $P_r(r_{n,m})$ of the ratios $r_{n,m}$ that are displayed in the lower panels of figure \ref{fig:PDFUU} do not display such widening. For sufficiently large $n$  all PDFs collapse to the same functional form that depends only on the choice of $\Delta \theta_c$.  This implies that while $U_{n,m}$ are not self-similar under scale transformations their ratios
$r_{n,m}$ are!

\begin{figure}[H]
\centering
\includegraphics[width=0.3 \textwidth]{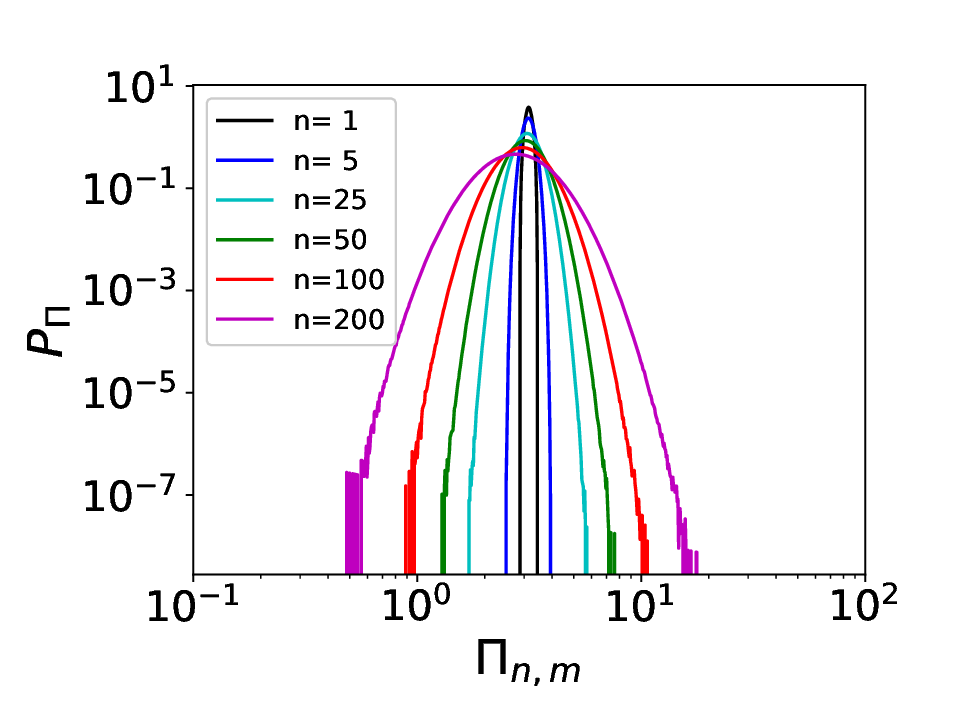}
\includegraphics[width=0.3 \textwidth]{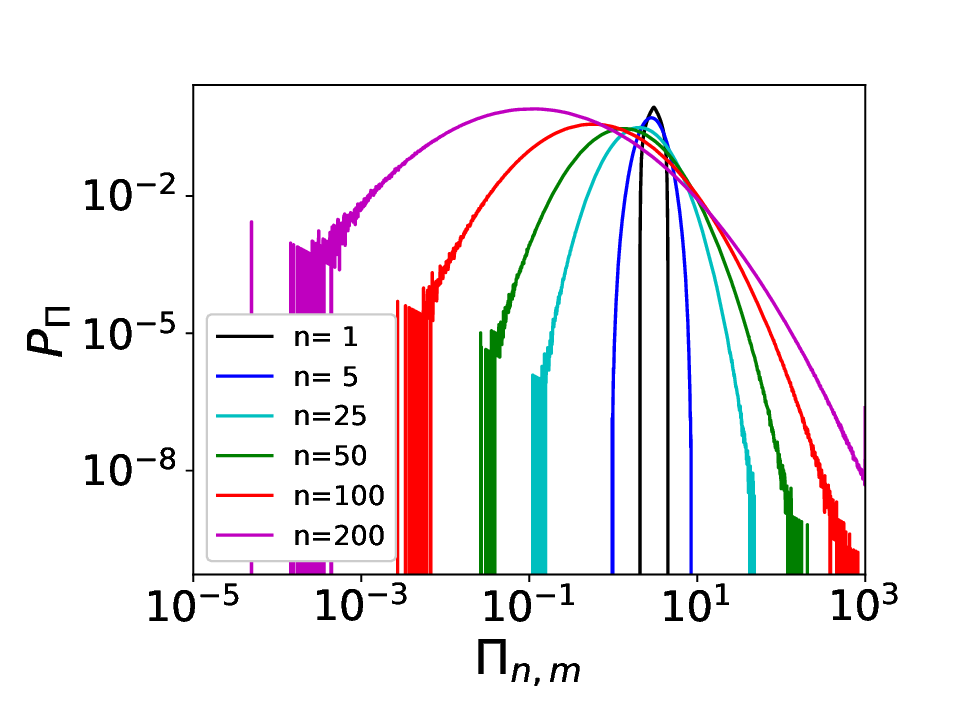}
\includegraphics[width=0.3 \textwidth]{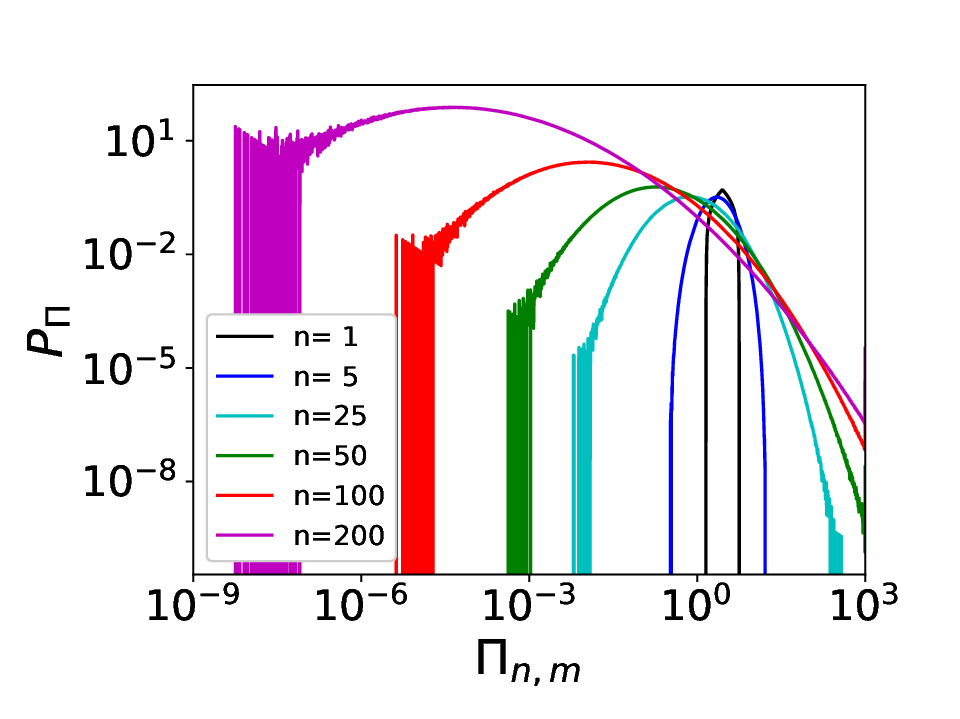}
\centering
\includegraphics[width=0.3 \textwidth]{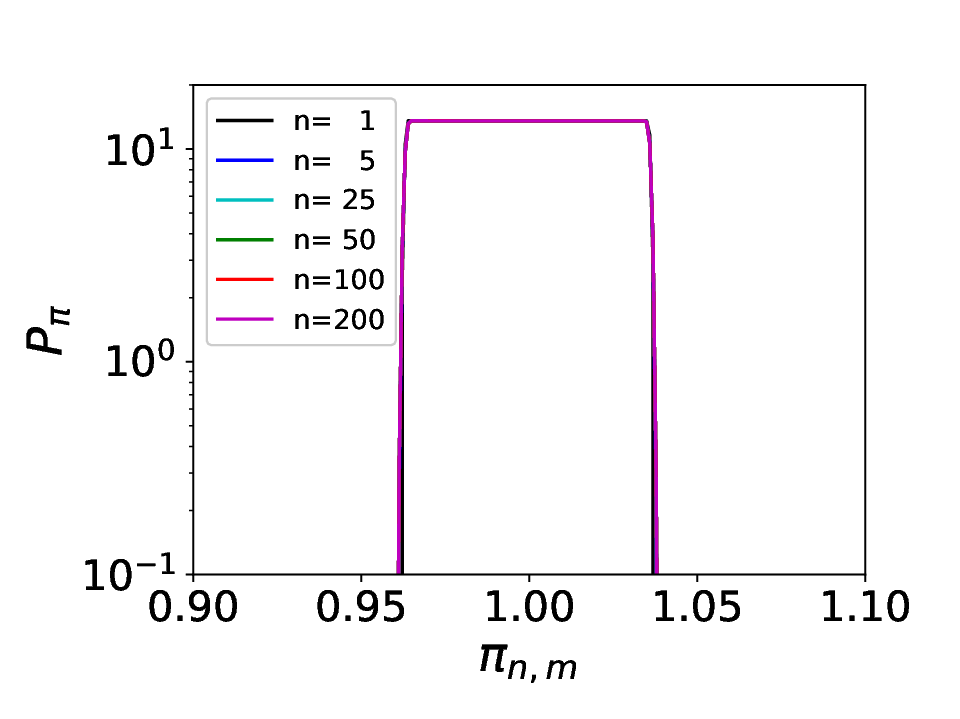}
\includegraphics[width=0.3 \textwidth]{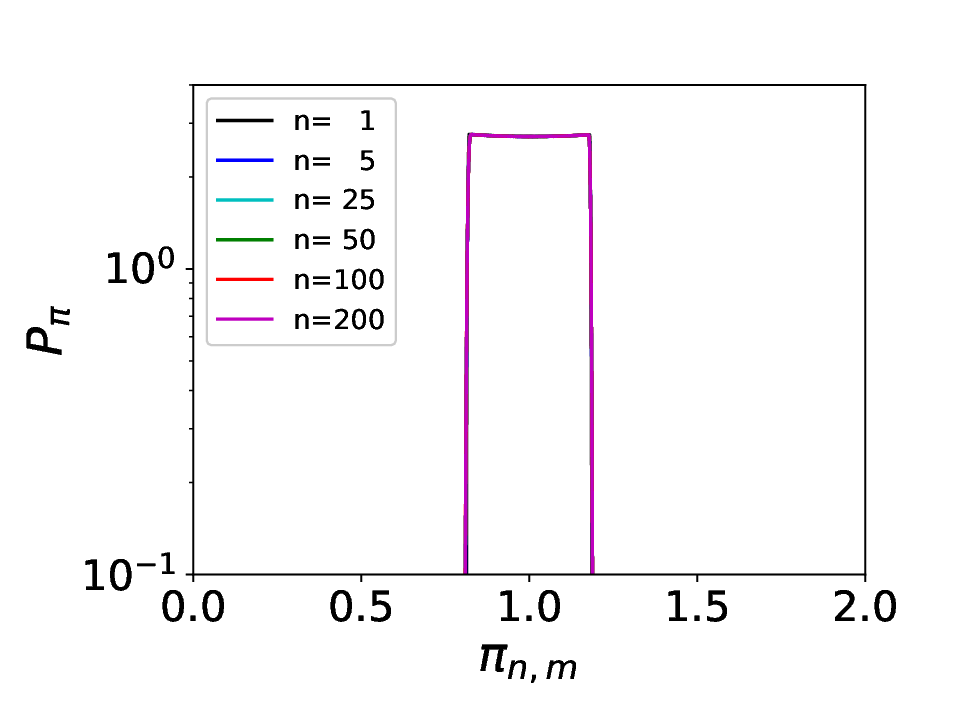}
\includegraphics[width=0.3 \textwidth]{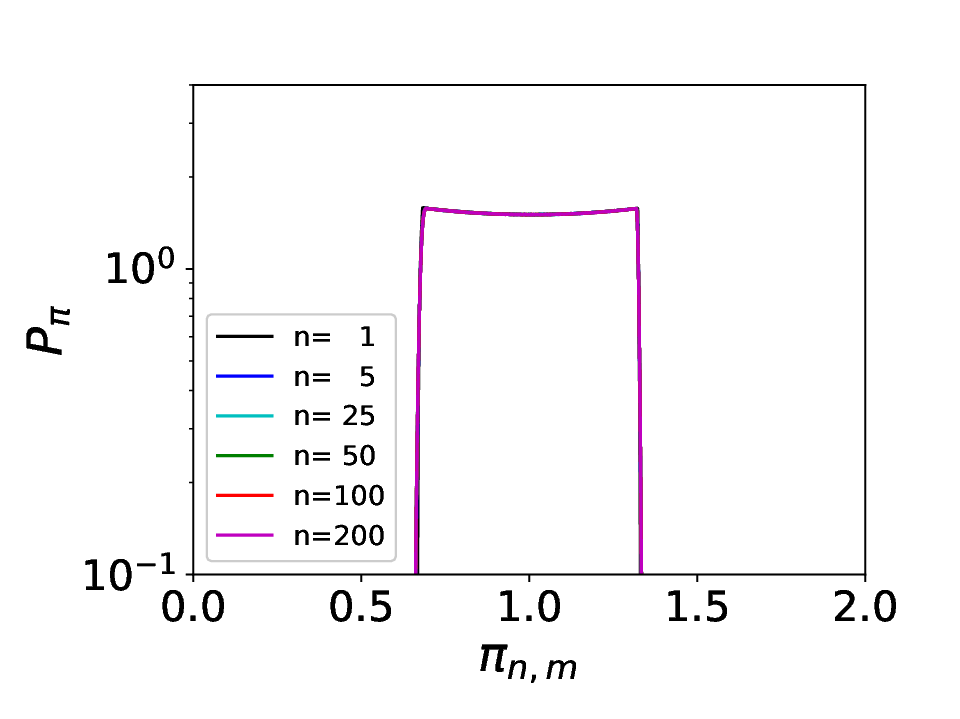}
\caption{{\bf Top panels:} PDFs $P_\Pi(\Pi_{n,m})$ of the fluxes $\Pi_{n,m}$ for the three different ensembles (left $\Delta \theta / \Delta \theta_c = 0.1$, center $\Delta \theta / \Delta \theta_c = 0.5 $ and right $\Delta \theta / \Delta \theta_c = 0.9$) for different values of $n$.
{\bf Bottom panels:} PDFs $P_\pi(\pi_{n,m})$ of the velocity ratios $\pi_{n,m}$ for the same ensembles and the same $n$. \label{fig:PDFFL}  }
\end{figure}   
The same behavior can be seen for the energy fluxes $\Pi_{n,m}$.  
In the top panels of figure \ref{fig:PDFFL} we plot the PDFs $P_\Pi$ of $\Pi_{n,m}$ for the same values of $\Delta \theta$ and $n$ as in figure \ref{fig:PDFUU}. As with the velocity amplitudes $U_{n,m}$ as $n$ is increased the PDFs of $\Pi_{n,m}$ widen without collapsing to to an $n$-independent form. In the lower panel of the same figure we plot the 
the PDFs $P_\pi(\pi_{n,m})$ of the flux ratio $\pi_{n,m}$. It is defined as
\begin{equation}
    \pi_{n,m} = \frac{\Pi_{n,m}}{\Pi_{n-1,m^*}}
\end{equation}
that after little algebra and using \ref{eq:flux} and \ref{eq:solx} leads to
\begin{equation}
    \pi_{n+1,m'} = 1 + f(r_{n,m})  \cos(2\theta_{n,m}) 
    \label{eq:flra}
\end{equation}
where $f(r)=1+(a/2c)^2r^2/(a/c+r)$. 
The flux ratio, much like the velocity ratio $r_{n,m}$, does converge to an $n$ independent PDF as
large values of $n$ are reached. Furthermore, the functional form of this PDF appears to be flat
limited by a minimum and a maximum value of $\pi_{n,m}$. This appears to be so because $f(r)$ in \ref{eq:flra}
varies little with $r$ for small variations of $r$ and the variations of $\pi_{n,m}$ are mostly controlled
by the variations of $\theta_{n,m}$.
%, with $f(r)\simeq1$ so that $\pi_{n,m}$ is limited by $1+\cos(2\theta_{min})<\pi_{n,m}<1+\cos(2\theta_{max})$.

%%%%%%%%%%%%%%%%%%%%%%%%%%%%%%%%%%%%%%%%%%%%%%%%%%%%%%%%%%%%%%%%%%%%%%%
\begin{figure}[H]                                                    %%
\centering
\includegraphics[width=0.3 \textwidth]{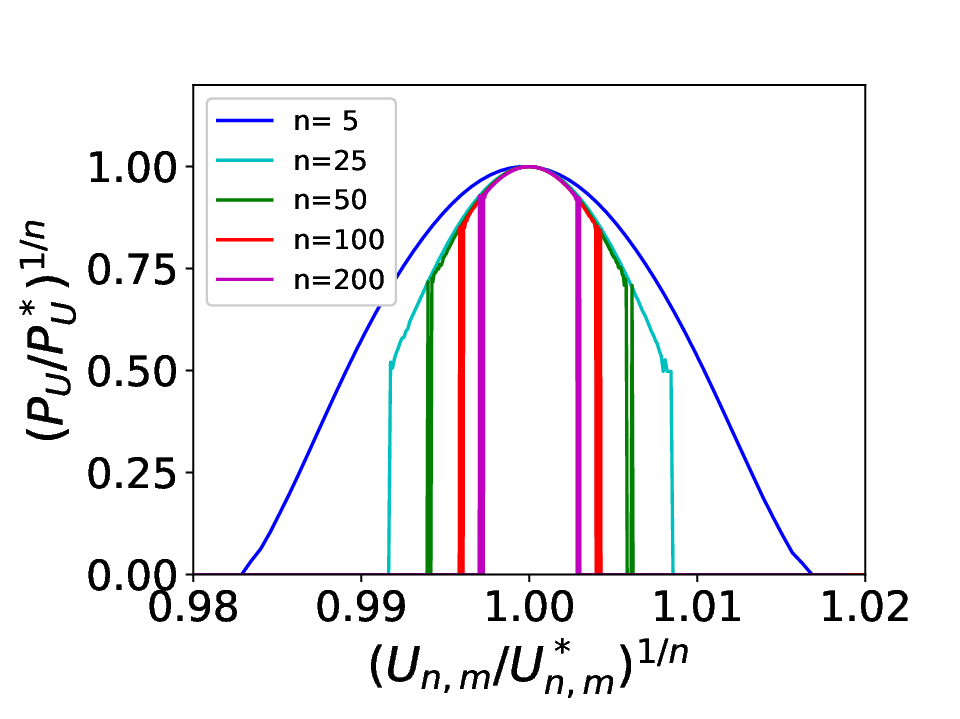}
\includegraphics[width=0.3 \textwidth]{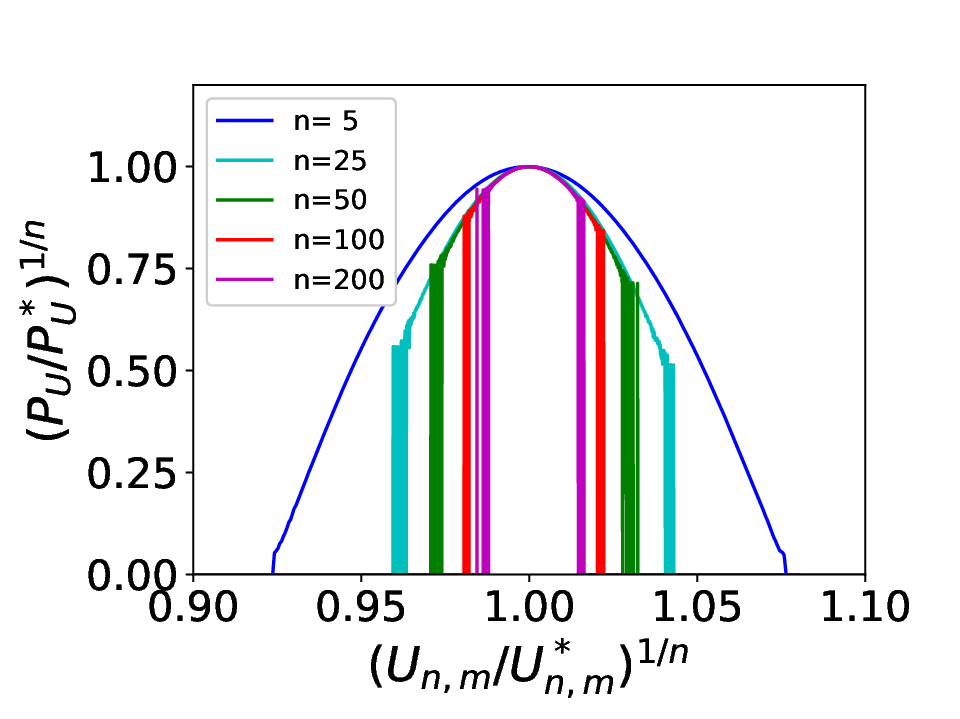}
\includegraphics[width=0.3 \textwidth]{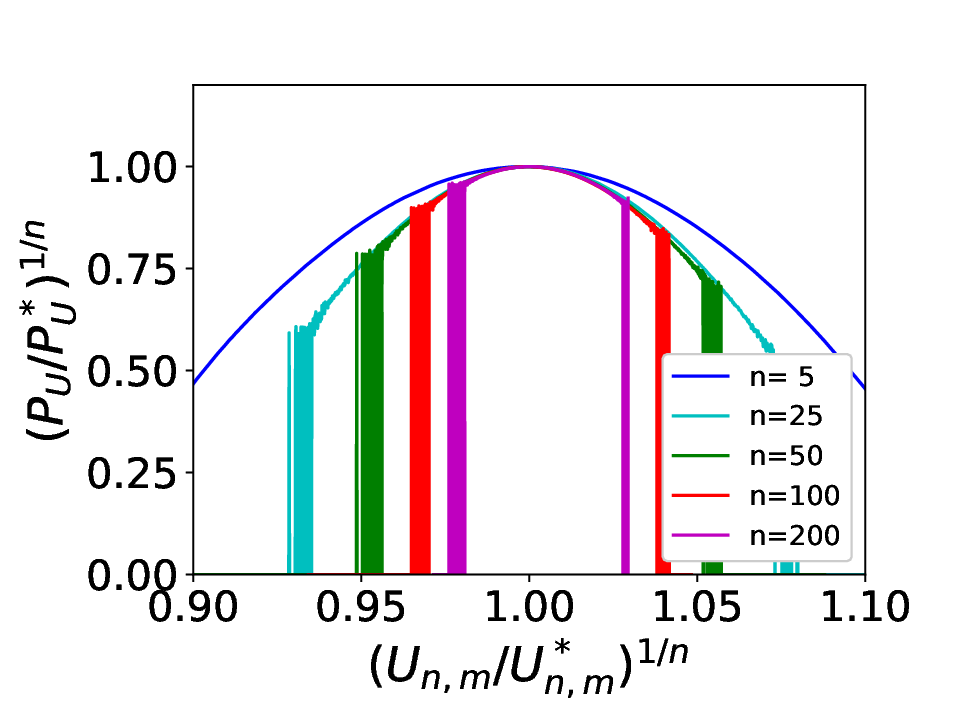}
\centering
\includegraphics[width=0.3 \textwidth]{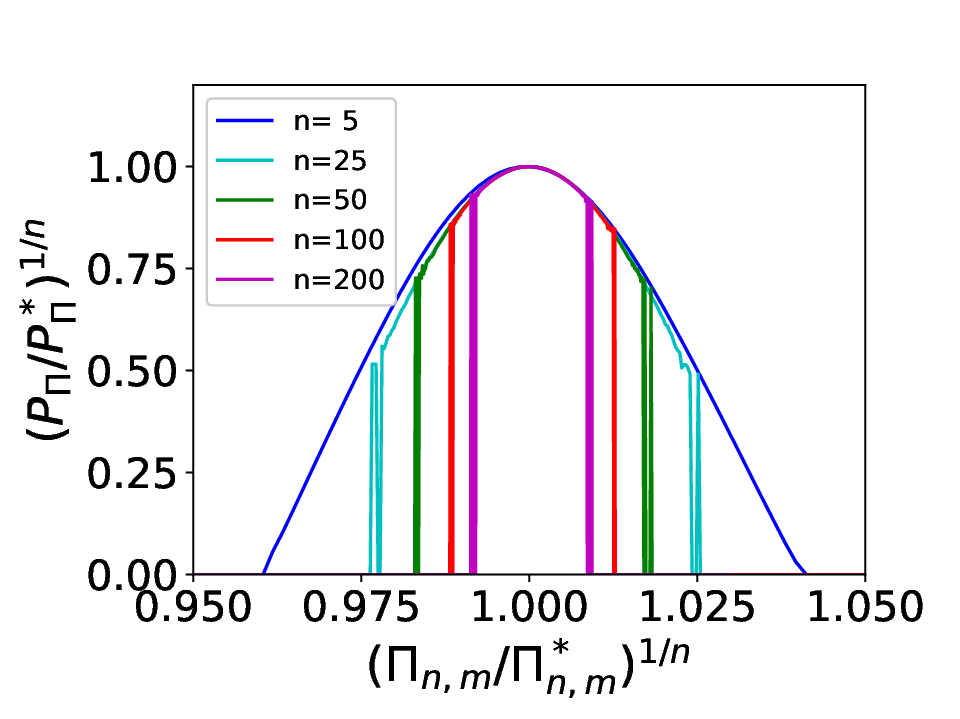}
\includegraphics[width=0.3 \textwidth]{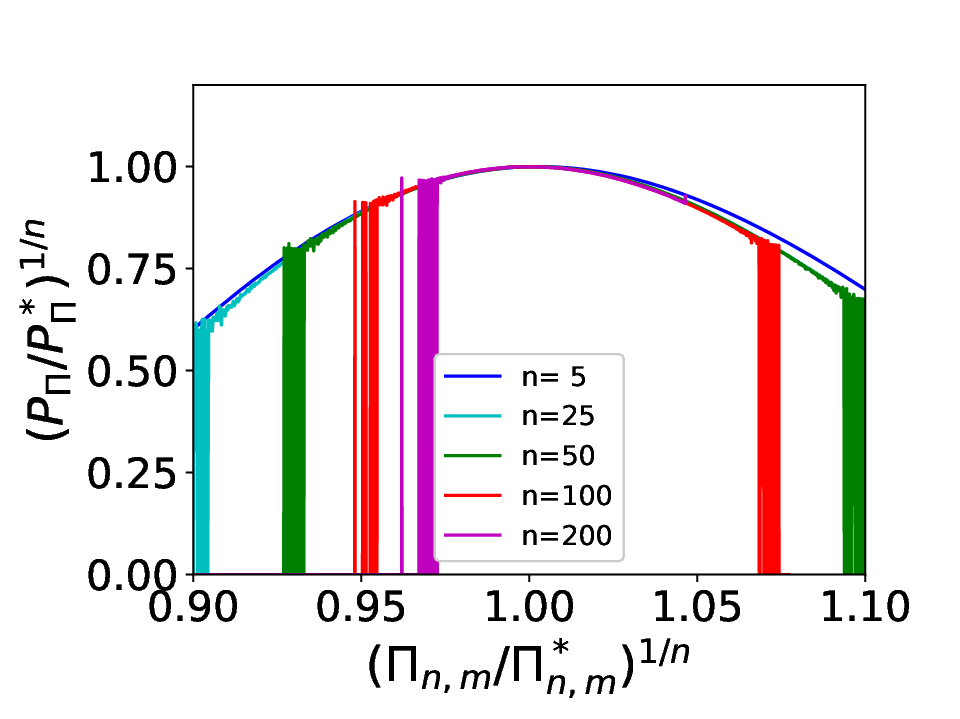}
\includegraphics[width=0.3 \textwidth]{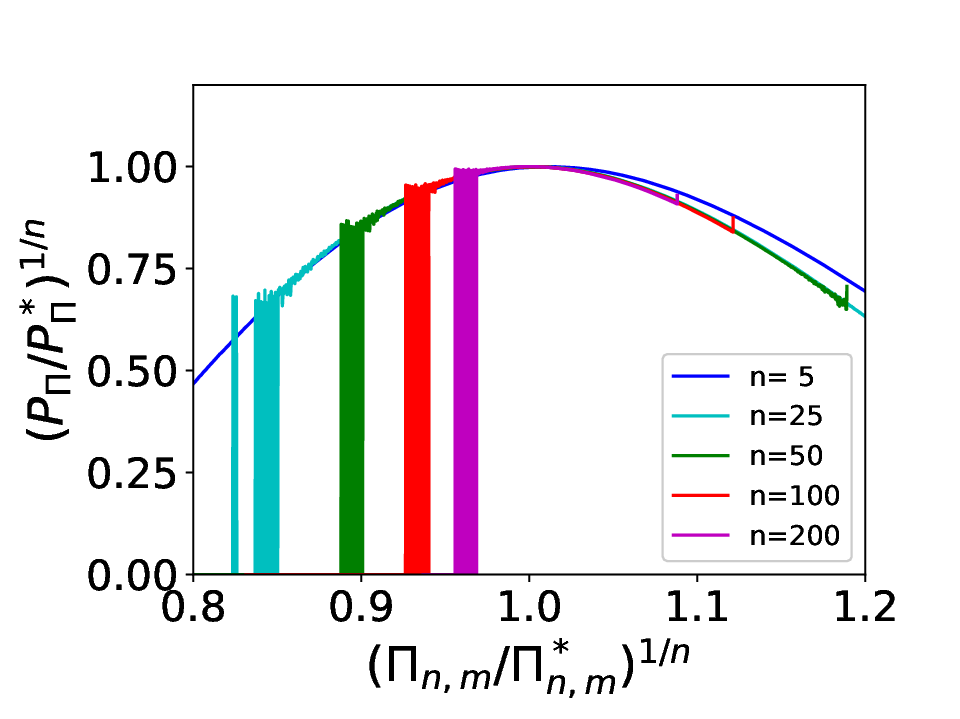}
\caption{{\bf Top panels:} PDFs $P_U(U_{n,m})$ for the different cases examined normalised
using the predictions of large deviation theory. (left $\Delta \theta / \Delta \theta_c = 0.1$, center $\Delta \theta / \Delta \theta_c = 0.5 $ and right $\Delta \theta / \Delta \theta_c = 0.9$)
{\bf Bottom panels:} The same for the PDFs $P_\Pi(\Pi_{n,m})$.
\label{fig:PDFRSCL}  }
\end{figure}   
%\unskip                                                             %%
%%%%%%%%%%%%%%%%%%%%%%%%%%%%%%%%%%%%%%%%%%%%%%%%%%%%%%%%%%%%%%%%%%%%%%%

Given that the PDFs of  $r_{n,m}$ and $\pi_{n,m}$ arrive at an $n$-independent form at large $n$
has some implications for the evolution in $n$ of the PDFs $P_U,P_\Pi$. Both $U_{n,m}$
and $\Pi_{n,m}$ can be written as a product of all $r_{n',m}$ and $\pi_{n',m}$ with $n'\le n$.
As a result the logarithms of $U_{n,m}$ and $\Pi_{n,m}$ can be written as
\begin{equation}
\ln\left(U_{n,m}\right)  = \ln\left(U_{1,1}\right) + n L_U, \quad
\ln\left(\Pi_{n,m}\right)  = \ln(\Pi_{1,1}) + n L_\Pi \label{eq:trsf}
\end{equation}
where $L_U$ and $L_\Pi$ stand for the mean value of the logarithms of $r_{n,m}$ and $\pi_{n,m}$ respectively:
\begin{equation}
L_U   = \frac{1}{n} \sum_{n'=1}^{n} \ln(r_{n',m}), \quad \mathrm{and} \quad
L_\Pi = \frac{1}{n}\sum_{n'=1}^{n} \ln(\pi_{n',m}).  
\end{equation}
The properties of $U_{n,m}$ and $\Pi_{n,m}$ remind the random cascades studied in the past 
\citep{novikov1964turbulent,yaglom1966effect,mandelbrot1999intermittent}. However while the random cascade models were not conserving energy in the present model energy is conserved exactly. 
An other important difference here is that $r_{n',m}$ and $\pi_{n,m}$ are not independent but each one depends on the value of the previous one. Nonetheless we can proceed assuming such independence
although not entirely correct. In that case $P_U$ and $P_\Pi$ can be reconstructed using large deviation theory
\cite{touchette2009large}.
In this framework $L_U$ and $L_\Pi$ follow for large $n$ a distribution of the form
\begin{equation}
P_{L_U}(L_U)    \propto \exp[-n I_U(L_U)], \quad \mathrm{and} \quad
P_{L_\Pi}(L_\Pi)  \propto \exp[-n I_\Pi(L_\Pi)] \label{eq:LDP}
\end{equation}
where $I_U$ and $I_\Pi$ are called the rate functions that can in principle be obtained from $P_r$ and $P_\pi$
using the Legendre-Fenchel transform \cite{touchette2009large}. Here we limit ourselves in noting that if $P_{L_U}$ and $P_{L_\Pi}$
follow the form of eq. \ref{eq:LDP} then the distribution of $U_{n,m}$ and $\Pi_{n,m}$ that are linked to  
$L_U$ and $L_\Pi$ by \ref{eq:trsf} should take the form
\begin{equation}
    P_U(U_{n,m})\propto \exp\left[ -n I_U\left( \frac{1}{n}\ln\left( \frac{U_{n,m}}{U_{1,1}} \right)  \right) \right] , \quad  P_\Pi(\Pi_{n,m}) \propto \exp\left[ -n I_\Pi\left( \frac{1}{n}\ln\left( \frac{\Pi_{n,m}}{\Pi_{1,1}} \right)  \right) \right] 
\end{equation}
where only the largest terms in $n$ are kept. To test this prediction we plot in figure \ref{fig:PDFRSCL}
$(P_U/P_U^*)^{1/n}$ as a function $(U_{n,m}/U^*)^{1/n}$ (top panels) and 
$(P_\Pi/P_\Pi^*)^{1/n}$ as a function $(\Pi_{n,m}/\Pi^*)^{1/n}$ where 
$U^*$ and $\Pi^*$ corresponds to the value the probability obtains its maximum $P_U^*,P_\Pi^*$.
With this normalization the PDFs both for $U_{n,m}$ and for $\Pi_{n,m}$ collapse,
indicating that the large deviation principle works well for this model.

\begin{figure}[H]
\includegraphics[width=0.3 \textwidth]{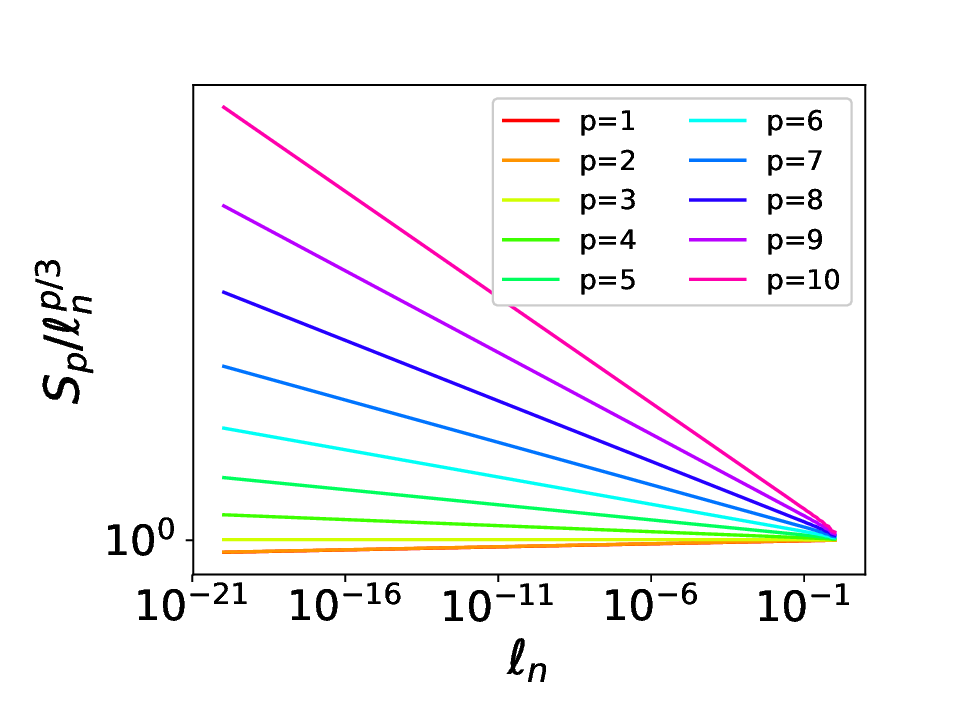}
\includegraphics[width=0.3 \textwidth]{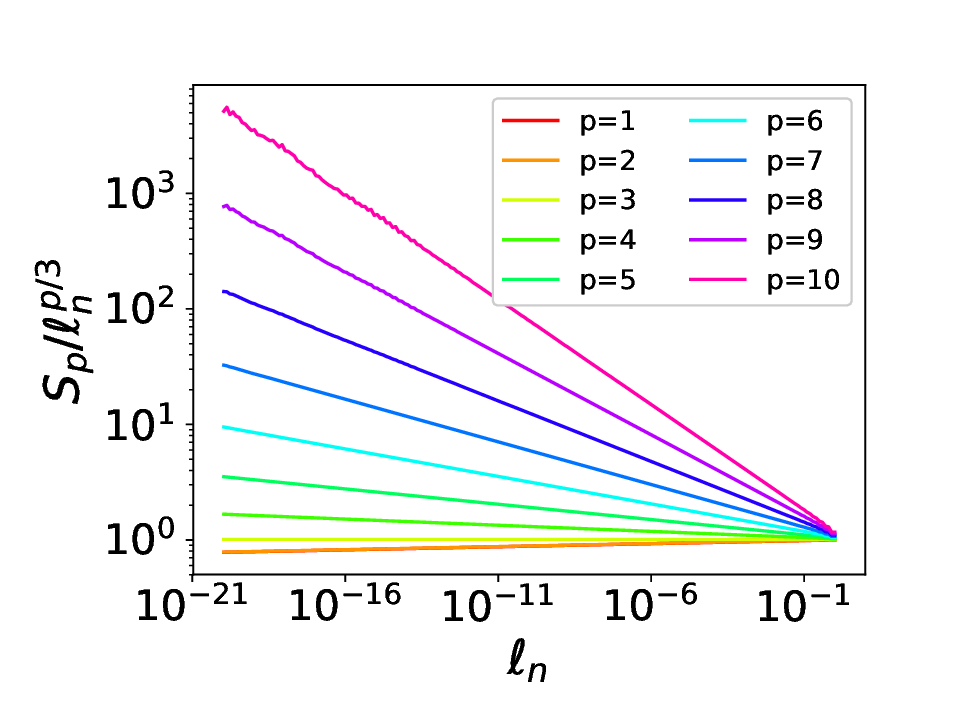}
\includegraphics[width=0.3 \textwidth]{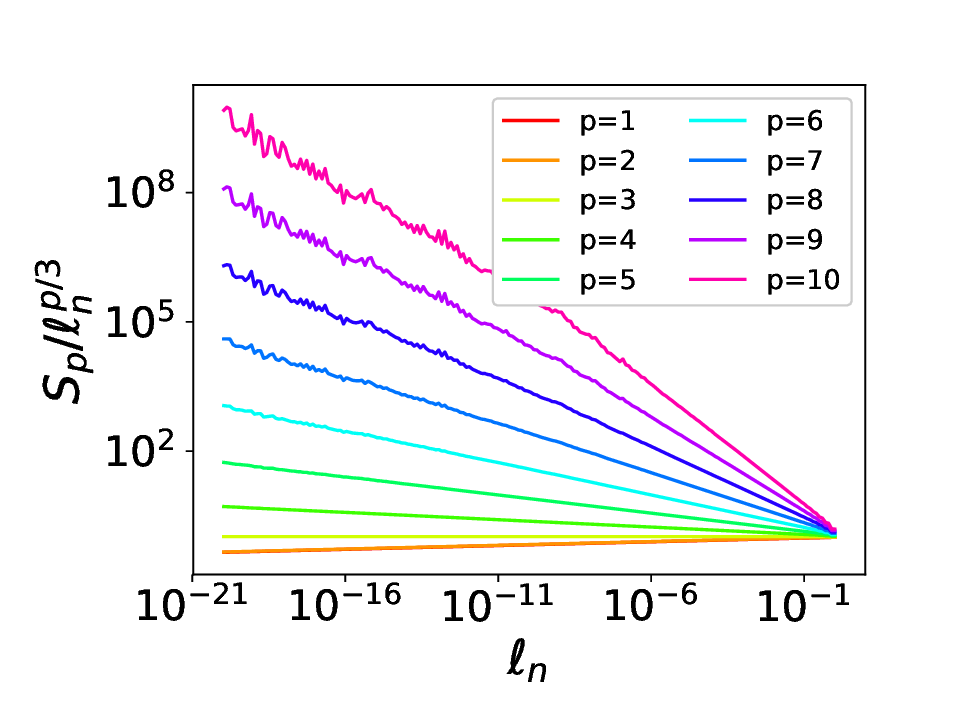}
\includegraphics[width=0.3 \textwidth]{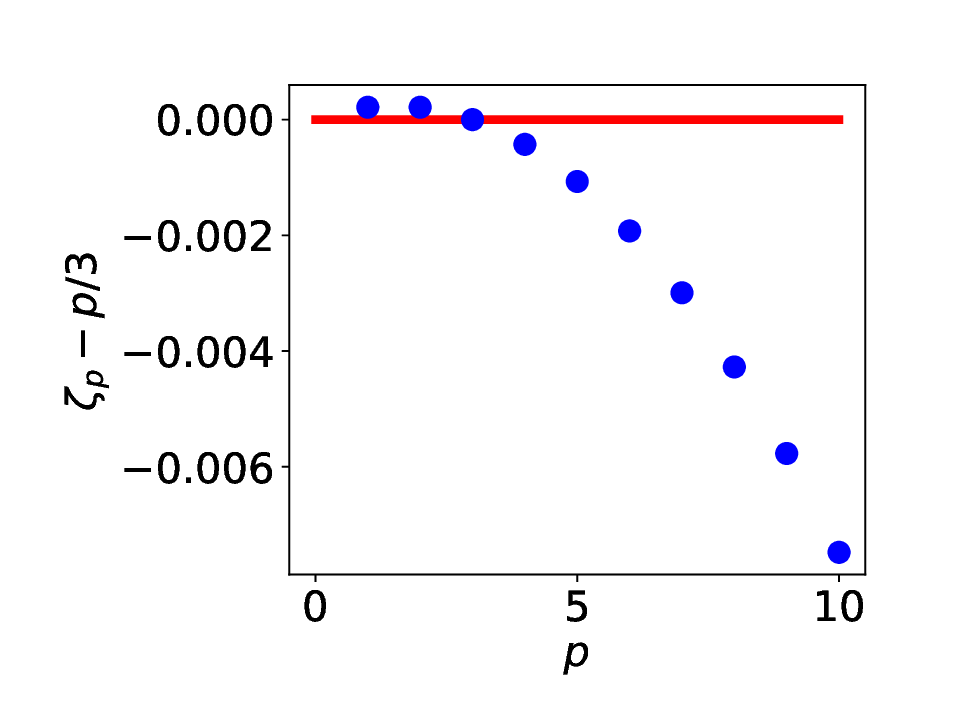}
\includegraphics[width=0.3 \textwidth]{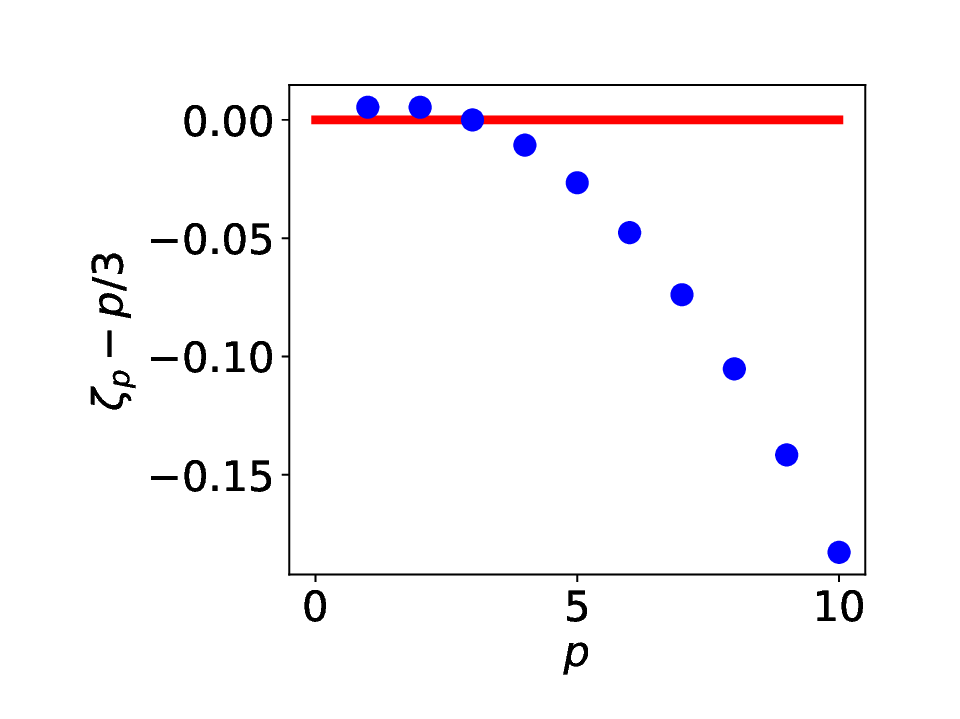}
\includegraphics[width=0.3 \textwidth]{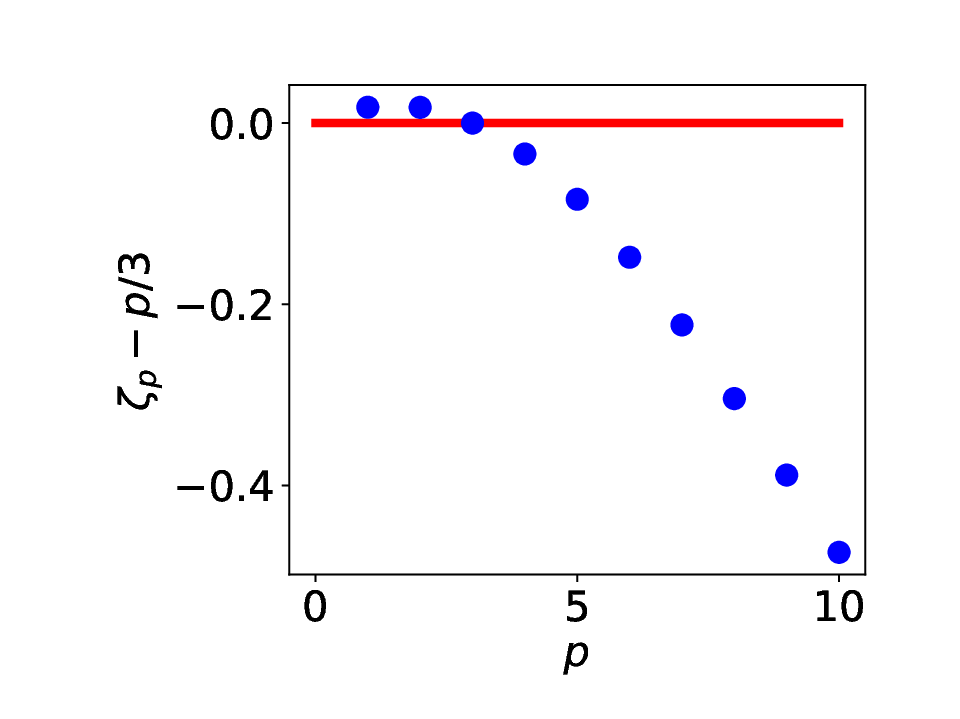}
\caption{{\bf Top panels:} Structure functions up to 10th order for the three
different values of $\Delta\theta$ examined.(left $\Delta \theta / \Delta \theta_c = 0.1$, center $\Delta \theta / \Delta \theta_c = 0.5 $ and right $\Delta \theta / \Delta \theta_c = 0.9$)
{\bf Bottom panels:} The resulting exponents $\zeta_p$. \label{fig:STRCT} \\ }
\end{figure}   
%\unskip
As a final look in the intermittency problem we display in the top panels of figure \ref{fig:STRCT}
the first ten structure functions $S_p(\ell_p)$ defined as
\begin{equation}
    S_p(\ell_n) = \left\langle U_{n,m}^p \right \rangle 
\end{equation}
where the angular brackets stand for ensemble average. 
The structure functions have been normalized by the Kolmogorov scaling to emphasise 
the differences. The structure functions are fitted to power-laws 
\begin{equation}
    S_p(\ell_n) \propto \ell_n^{\zeta_p}
\end{equation}
and the measured exponents $\zeta_p$ are plotted in the lower panels of figure
\ref{fig:STRCT}. The exponents show similar behavior with real turbulence displaying 
larger values for $p<3$ and smaller values for $p>3$ while the exact result $\zeta_3=1$
is satisfied. It is worth noting that the deviations from the Kolmogorov scaling 
are not universal but depend on our choice of ensemble that is controlled by  $\Delta \theta$. 

%%%%%%%%%%%%%%%%%%%%%%%%%%%%%%%%%%%%%%%%%%
%\section{Materials and Methods}

%%%%%%%%%%%%%%%%%%%%%%%%%%%%%%%%%%%%%%%%%%
%\section{Results}

%%%%%%%%%%%%%%%%%%%%%%%%%%%%%%%%%%%%%%%%%%
\section{Discussion and conclusion}

One can argue that the exact stationary solutions obtained in this work little do they have to do with
real turbulence that displays  chaotic spatio-temporal dynamics. This maybe true and multy branch models
with two neighbour interactions as in \cite{aurell1994hierarchical,aurell1997binary} that display chaotic dynamics 
should be further investigated. The present results however do point to
a clear instructive demonstration of how intermittency can appear in realistic flows and how it can be modeled.  
Furthermore, it leads to a series of clear messages which are described bellow that are of great use in 
future turbulence research and can guide measurements in numerical simulations and experiments.

First, we note that intermittency appearing in stationary fields found here comes in contrast with the typical 
shell model studies in single branch models for which intermittency comes from the temporal dynamics alone
as only a single structure exists for each scale $\ell_n$. In the latter case intermittency has been linked to 
the temporal dynamics through the fluctuation dissipation theorem \cite{aumaitre2023}. In reality, both temporal and spatial 
dynamics contribute to the presence of intermittency and their role needs to be clarified.

In the present model randomness comes from our choice of $\theta_{n,m}$ and the resulting intermittency depends on that choice. In reality, (or in more complex shell models) such randomness will come from local chaotic dynamics that need to 
studied in order to clarify which processes lead to enhanced cascade and with what probability.

Perhaps, the most interesting implication of this work is that it suggests new ways to plot data from 
experiment and numerical simulations. 
One way suggested by this work is instead of focusing on the PDFs of  velocity differences
experimental or numerical data could focus on the PDFs of ratios of velocity differences.
The latter are shown in this work to become scale independent and could lead to more precise measurements. 
An alternative way is to re-scale the PDFs of velocities differences using the large deviation prediction
\ref{eq:LDP} as was done in figure \ref{fig:PDFRSCL}. Of course in realistic data $n\propto \ln(L/\ell_n)$
is not precisely defined and an optimal choice should be searched for.

A good model of a complex phenomenon, to the authors opinion, is not one that  quantitatively reproduces experimental measurements through parameter fitting but rather one that unravels the processes involved. 
To that respect we believe that the present model and results are very fruitful. We only hope that this work would come close to the standards set by Jack Herring. AA met Jack Herring during his ASP post doc in 2004-2006. Jack is fondly remembered stopping by the offices of post-docs just to see if they are OK. He will be greatly missed.

%%%%%%%%%%%%%%%%%%%%%%%%%%%%%%%%%%%%%%%%%%
%\section{Conclusions}
%\vspace{6pt} 

%\funding{
This work was also supported by the Agence nationale de la recherche (ANR DYSTURB project No. ANR-17-CE30-0004).
%}

%\acknowledgments{
 AA would also like to acknowledge the help of the Pétrélis brothers, Francois and Nicolas,
for their help with the large deviation theory. We would also like to thank Annick Pouquet for 
inviting us to write an article in this special issue dedicated to Jack Herring.
%}

\bibliography{Atmosphere.bib}

\end{document}